\documentclass[CRPHYS,Unicode,manuscript]{cedram}
\newcommand*\fdeg{\hbox{$^\circ$~}}

\title{North Polar Spur/Loop I: gigantic outskirt of the Northern {\it Fermi bubble} or nearby hot gas cavity blown by supernovae?}

\author{\firstname{Rosine} \lastname{Lallement}\IsCorresp}
\address{GEPI/Observatoire de Paris,PSL*, 5 pl. J. Janssen, 92195 Meudon, France}
\email[R. Lallement]{rosine.lallement@obspm.fr}


\begin{abstract}
Radio continuum, microwave and gamma-ray images of the Milky Way reveal a conspicuous, loop-like structure that fills almost half of the northern Galactic hemisphere, called Loop I. The interior of Loop I is the most conspicuous region shining in soft X-rays, whose eastern base is a remarkably bright, elongated structure seeming to emerge from the Galactic plane, dubbed the North Polar Spur (NPS). After 40~years of debates, two very different, contradictory views of Loop I/NPS are still defended: on the one hand, the NPS is a gigantic volume of expanding hot gas that envelops and extends the northern ``Fermi Bubble'' (FB) known to be blown by the Galactic center, and Loop I marks the shock front; on the other hand, the NPS is totally independent of the northern FB, it is a nearby, ordinary cavity of hot gas blown by supernovae, Loop I is its shock front and both are coincidentally located in the direction of the FB. To an observer at the Sun, both can produce the same perspective view, although the former has a size comparable to the Milky Way itself, and the latter a diameter of a few hundreds parsecs. The energy involved varies by 3-4~orders of magnitude, and the solution has important consequences on the structure and history of our Galactic neighborhood, on the age of the North and South FBs and the activity at the Galactic center. Moreover, whatever are the actual shape and distance of Loop I/NPS, accurate modeling of the polarized emission associated with Loop I is important for CMB foreground removal. After a short review, I discuss recent results which have a connection with Loop I/NPS. Some of them have been used as arguments in the two opposite ways, while for others the connections with LoopI/NPS have been overlooked. They involve very different spacecraft, from a 12~Kg Cubesat (HaloSat) to major space-borne observatories (HST, Gaia, and Spektr-RG). I make use of updated 3D~maps of dust and a recent massive star catalog. I distinguish arguments based on geometric similarity or dissimilarity from those derived from measurements and physical models. Considering all past and recent constraints, it is clear that there is no entirely local or entirely distant scenario that is free from contradictions with some of the data analyses or from improbable coincidences. I discuss a speculative scenario, partially inspired by recent BF and Milky Way gas models, combining near and far aspects and seeming to be able to meet the various constraints. However, new data and models are needed to bring the controversy to a close and we can still expect new episodes of this long story.
\end{abstract}

\begin{document}
\maketitle

\section{Introduction}

\begin{figure}[tbp]
\includegraphics[width=0.52\linewidth]{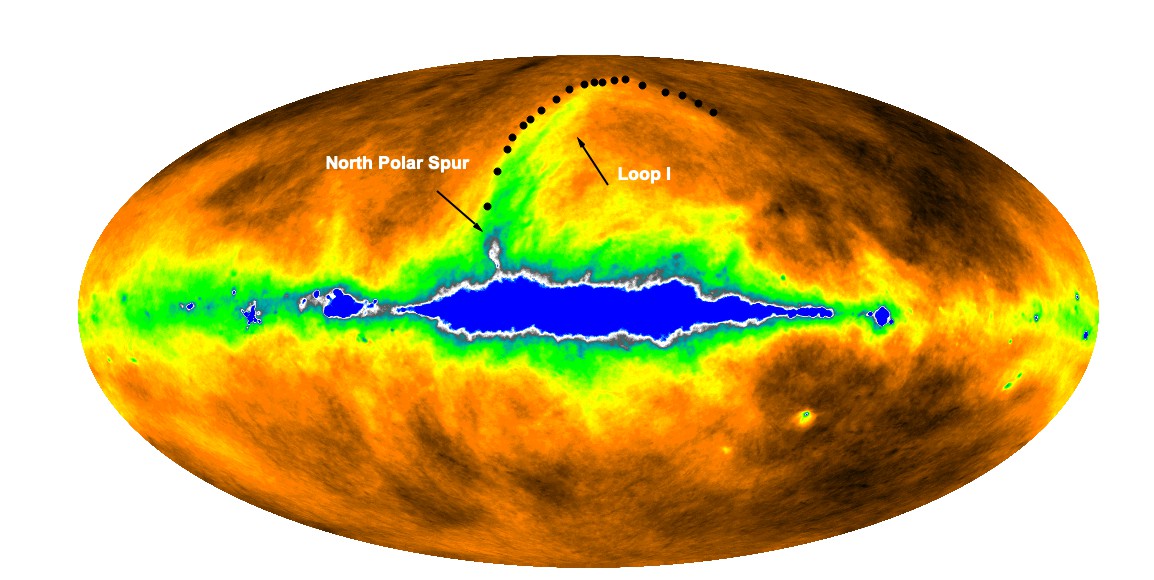}
\includegraphics[width=0.47\linewidth,height=3.35cm]{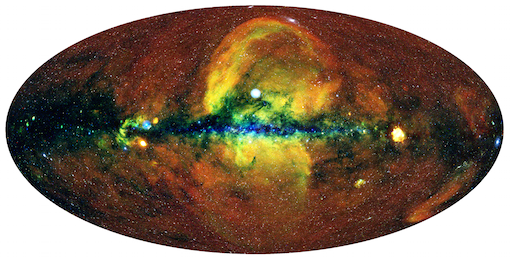}
\caption{408~MHz radio continuum (from~\cite{Haslam74}, reprocessed by~\cite{Remazeilles15}) and Spektr-RG/eROSITA soft X-rays (from~\cite{Predehl20}) all-sky maps revealing radio Loop 1 in the northern hemisphere and its associated X-ray bright area, in particular the eastern low latitude part of the latter, the NPS. The X-ray image is a composite of three bands: red(0.3-0.6~keV), green(0.6-1.0~keV) and blue(1.0-2.3~keV). Both maps are in Aitoff Galactic coordinates. North is at top, East is left of center. The black signs in the 408~MHz map are also reported in the ROSAT X-ray maps used in Section~\ref{distfromxray}.}\label{fig:presentatx}
\end{figure}

Loop I is a spectacular $\simeq 70$\fdeg x 50\fdeg loop-like feature shining above the Galactic Center at radio continuum, microwave, and more weakly gamma-ray wavelengths. Its contour delimitates a $\simeq 1$-2~keV X-ray bright area which is characterized by a strong brightness enhancement of it eastern low latitude part, the North Polar Spur (NPS). Among the long-lasting controversies in astrophysics, the 40-year-old debate about the physical size, the location and the origin of the NPS/Loop I association (hereafter LNPS) is one of the most surprising. As a matter of fact, here the scene is not the distant Universe, instead it is our Milky Way, i.e. a well-observed astrophysical object. Moreover, the structure in debate is not a faint or minor feature, instead it is a major, conspicuous object. Last but not least, the two debated interpretations are so radically different that it is hard to believe that they both co-existed for so long.

After an introduction to the observations and debate, a series of recent results related directly or indirectly to the origin of the LPNS are presented, along with their implications. Although all arguments influence the entire LNPS structure and are often interconnected, they are separated into three categories related to the overall geometry, a fourth section devoted to inferences from X-ray data, and finally a section on stellar and interstellar constraints on the nearby super-bubble hypothesis. The final Section~\ref{sect7} contains a summary and discussion of potential solutions and future measurements that could definitely close the debate.

\subsection{First observations}

\begin{figure}[!htbp]
\includegraphics[width=0.99\linewidth]{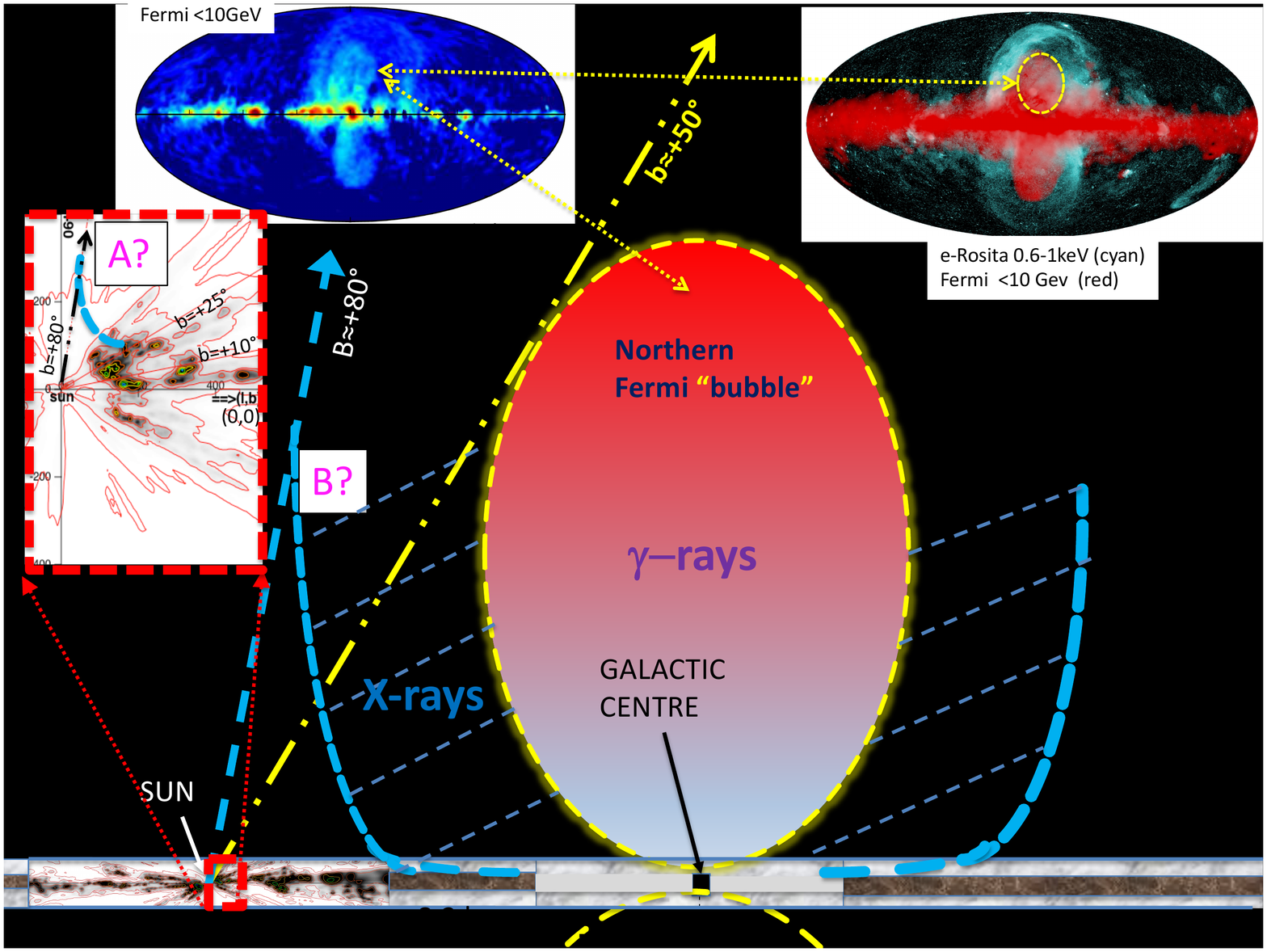}
\caption{\textbf{Top left:} Gamma-ray sky (from~\cite{Casandjian15}). The FBs correspond to the double-lobed structure centered exactly on the GC and reaching up to b$\simeq 50$-55\fdeg above and below the plane. A fainter, wider loop similar in location to Loop I is also visible in the north, surrounding the FB bubble and reaching b$\simeq +80$\fdeg\!\!. \textbf{Top right}: Soft X-ray sky measured by e-Rosita (cyan), with the Fermi image superimposed (in red), and showing the respective locations of the X-ray bright areas and the FBs in both hemispheres (from~\cite{Predehl20}). The southern X-ray structure is much fainter than its northern counterpart, however, in both hemispheres the X-ray bright regions encircle the FBs in a similar way. The northern, fainter $\gamma$ray loop is not visible in the superimposed image, however it can be seen that it would encircle closely the X-rays. The external regions of the X-ray extended area correspond quite closely to the radio loop shown in Figure~\ref{fig:presentatx}. \\
\textbf{Main figure and illustration of the two interpretations:} Schematic view of the meridian plane perpendicular to the Galactic plane and containing the Sun and the GC. The Milky Way thin disk of dust and gas is shown as a 800~pc wide dark and light grey band extending about 15~kpc on both sides of the GC, except for the region internal to the 3~kpc molecular ring around the GC white band, and for the 6~kpc wide region around the Sun where we use the dust distribution (from~\cite{Vergely22}). Assuming that the FBs are symmetric around an axis perpendicular to the Plane, the northern FB would occupy the region within the dashed yellow ellipse. If LNPS X-rays are associated to the FBs and surrounding them, the X-ray emitting regions could be bounded by the dashed blue line B (see text). This corresponds to the interpretation of NPS/Loop I as the envelop and outskirts of FB north. A zoom showing the dust distribution within $\pm 3$~kpc from the Sun is shown at left (dashed red-line rectangle, from~\cite{Vergely22}). Units are parsecs, the Sun is at (0,0), the GC direction is to the right and the north Galactic pole direction to the top. In the local interpretation of LNPS, X-ray, radio and microwave emissions would come from a very close region seen from low to high latitudes, whose periphery is schematized as arc A (blue dashed line in the zoom).}\label{fig:figscheme}
\end{figure}

Observations directly or indirectly related to LNPS could fill long chapters, and we restrict this introduction to early multi-wavelength observations and to the main findings of the last decade. Several presentations of Loop I and discussions about its origin have been published, see the very deep analysis by~\cite{Planck16} and references therein, and for a very recent discussion see~\cite{Panopoulou21}. Loop I was first detected in radio as the most prominent Galactic loop, with the NPS at its bottom\linebreak(\!\!\cite{Large62, Berkhuijsen71, Haslam74, SofueReich79}). The two structures were later found to bound the most prominent diffuse feature of the entire sky in soft X-rays (\!\!\cite{Snowden97}), along with a wide region in the general direction of the Galactic center, the X-ray bulge. To illustrate the similarity between radio and soft X-rays, Figure~\ref{fig:presentatx} shows the 408~MHz continuum full-sky map (from~\cite{Haslam74}, reprocessed by~\cite{Remazeilles15}) and the very recent Specktr-RG/eROSITA composite soft X-rays (0.3-2.0~keV) full-sky image (\!\!\cite{Predehl20}). The contour drawn in the radio map on top of Loop I is superimposed on the ROSAT maps of~\cite{Snowden97} used in Section~\ref{distfromxray} and available numerically. Unambiguously, radio and X-rays trace the same object. Both NPS and Loop I were also mapped from ground in polarized radio continuum (\!\!\cite{Bingham67, Berkhuijsen71, Brouw76}) and later from space in total-intensity and polarized microwave emission during missions devoted to the CMB. More recently, a counterpart in gamma-rays was also found by~\cite{Casandjian09}, based on Fermi-LAT data (see Figure~\ref{fig:figscheme}).

After its discovery, it was hypothesized that LNPS is associated with a nearby hot gas cavity blown by a recent supernova (SN) or a series of supernovae. According to the classical scheme, soft X-rays are thermally emitted by million degree hot gas blown by the massive precursor stellar winds and SN ejecta to form a large remnant cavity, while the radio signal is due to synchrotron emission by supra-thermal electrons accelerated in external, shocked regions of the SN remnant (SNR). Indeed, spectra slopes of radio data were found compatible with the synchrotron mechanism. The proximity of the LNPS source was considered as inevitable. Due to its exceptionally large angular size, maintaining a physical dimension of LNPS in a realistic range implies a short distance, and, in addition, below Loop I are a series of hot young O, B stars associations, in particular the Scorpius-Centaurus group at $\simeq 130$~pc, ideally located above the Plane and a very likely source of recent supernovae. A first potentially contradicting observation is the absence of high velocity expanding shell at the periphery of LNPS, because million degree gas implies shock speeds on the order of tens to hundreds of km/s. This was explained by a two- or multiple-stage formation of the X-ray emitting volume associating an initial old (few million years) SNR episode and a recent re-heating of the blown cavity due to more recent explosions (see Section~\ref{Reheated}). The Local Cavity surrounding the Sun (often called Local Bubble, LB) and filled with hot, highly ionized gas is also generally believed to be a relic of similar events. An additional argument in favor of a local origin of LNPS was also very early proposed, namely the presence of high Galactic latitude HI filaments seeming to envelop Loop I (see, e.g., \cite{Berkhuijsen71}). All along the following decades, owing to increasingly detailed and sensitive gas and dust emission surveys, it was confirmed that there are nearby concentrations of gas and dust and especially high latitude arches in the same longitude-latitude range than the LPNS. Although no clear 3D view of their association with the large SNR episode responsible of LNPS was established, it was implicitly assumed that they are debris of a large shell formed around the hot gas (but see below).

In parallel, a very different interpretation of the LNPS started to be defended. In an attempt to explain the wide empty cavity (3~kpc radius) around the Central Molecular Zone (CMZ) at the Galactic center (GC), and the surrounding expanding ring of dense interstellar matter, it was proposed in 1977 that both were due to explosive events at the GC (\!\!\cite{Sofue77}), and, as part of the same model, that the propagation in the halo of the shock generated by the latest event has created huge volumes of hot gas and accelerated particles at the shock front. In the northern hemisphere the hot gas and the shock front are seen from the Sun vantage point as the LNPS. A series of models were since then regularly published by Sofue and co-workers, based on all new relevant observations,  including, from 2010, those of the Fermi Bubbles (FBs) discovered at that time and unanimously recognized as generated at the GC (see below). However, the local origin of LNPS and a total absence of link between LNPS and the FBs remained the preferred scenario for most astrophysicists.

\subsection{The Fermi \emph{bubbles} discovery}

A turning point occurred at the time of the successive discoveries of (i) the so-called WMAP \emph{haze} around the Galactic center, (ii) its Fermi gamma-ray counterpart and, finally, (iii) the ``Fermi bubbles''. Shortly, after the WMAP mission devoted to the CMB mapped the entire sky in several microwave frequency bands, \cite{Finkbeiner04} noticed a $\simeq$ 70\fdeg wide brightness enhancement around the GC area. The initially proposed origin of the ``haze'' was dark matter annihilation, but later analyses favored instead synchrotron emission by energetic electrons (\!\!\cite{Dobler08}). Subsequently, \cite{Dobler10} discovered the gamma-ray counterpart of the haze and suggested that the same population of relativistic electrons produces both microwave synchrotron emission and gamma rays through inverse Compton with the interstellar radiation field. Soon after, a reanalysis of the Fermi data by~\cite{Su10} revealed two large gamma-ray bubbles, extending 50\fdeg above and below the Galactic center, with a width of about 35\fdeg in longitude, the spectacular Fermi bubbles (FBs). Without any doubt, such giant volumes of highly energized gas with sizes comparable to the one of the Milky Way, are due to bursting activity at the Galactic center.

The existence and observed geometry of the northern FB, and the fact that it appears encircled by the LNPS (see Section~\ref{northsouth}) gave more weight to Sofue's hypothesis, and several models were developed to reproduce these gigantic, $\simeq 10$~kpc high structures, and study links between FBs and LNPS (\!\!\cite{Crocker15, Sarkar15, Sarkar19}). The detailed mechanisms at the origin of the FBs are currently lively debated. A series of X-ray, radio and IR data recently confirmed the existence of powerful jet-type outflows from the central parsecs at the GC, able to feed FB-type structures and suggesting that today's FBs are the result of a particularly strong activity several Myrs ago (see e.g.~\cite{Ponti21, Cecil21}).

\subsection{Illustration of the two interpretations}

Today, the debate about the size of the LNPS and its potential link with GC events and FBs is still very lively, as shown by very recent publications in favor of the two different interpretations (see below). Figure~\ref{fig:figscheme} schematizes the two interpretations and illustrates the strong difference in size scales. According to the $\simeq 50$\fdeg latitude of the northern FB top and the FB shape, \emph{its} height and maximal width reached at $\simeq2$~kpc altitude are on the order of 8.5 and 6~kpc respectively, if one assumes that the FBs are symmetric around the polar axes. In the GC scenario for the LNPS, the shocked halo gas envelops the FBs and, in the north, the high latitude limit of the X-rays corresponds to the highest latitude tangential direction to the X-ray boundary (schematized as arc B). To illustrate the local origin of the LNPS, an image of the dust distribution in the meridian plane within $\pm 3$~kpc from the Sun along the Sun-GC axis and within $\pm 400$~pc from the Plane towards the poles is inserted into the sketch, and a zoom on this image in the vicinity of the Sun is added. The zoom is used to illustrate the location and shape of a \emph{hypothetical shell generated by nearby SNRs} (arc~A) and viewed from the Sun within the same latitude range than \emph{the arc~B in the GC scenario}.

\subsection{On the link between NPS/Loop I and interstellar high-latitude dust and gas}\label{linkgasxray}

It has been often suggested that the high latitude HI and dust filamentary features between $\rm l=250$\fdeg and $\rm l=60$\fdeg are connected with the LNPS, based on morphological similitude between their location and orientation along the periphery of the X-ray emitting region, and this has been used as an argument in favor of the local hypothesis. Gas and dust filaments would mark the outer boundary of a shell surrounding the volume of heated gas. If proven, the existence of such an association would definitely close the controversy and favor a local LNPS. As a matter of fact, these high latitude dust or gas structures are all located at short distances, as demonstrated by various works on gas absorption~\cite{Puspitarini12}, stellar polarization (e.g., \cite{Panopoulou21}, and 3D extinction maps (\!\!\cite{Vergely22}, also see the $\rm b\simeq +60$\fdeg dust cloud in Figure~\ref{fig:figscheme} or other high latitude nearby (100-200~pc) faint structures in Figure~\ref{fig:loops_cuts_new} (right)). However, the association between gas and dust on one side, and X-ray, gamma-ray and synchrotron structures on the other side, has never been proven. A careful look at the various maps shows that there is no close correspondency (see, e.g., \cite[Figure~21]{Planck16}) and, as mentioned above, HI 21~cm spectra show that the gas is not expanding. None of the new sources of information discussed below involves directly the high latitude filaments of gas and dust, and we come back to them and their potential links with LNPS in the last section only.

\section{The North-South asymmetry}\label{northsouth}

One of the arguments against a GC origin of the LNPS is the absence of its southern counterpart. Because the FBs are nearly symmetric around the GC, one would expect a conspicuous X-ray-bright wide area around the southern FB, bordered by a South Polar Spur and a southern wide radio loop, similar to LNPS in the north. In the local source hypothesis, one may instead expect a non-symmetric prolongation to the south of X-ray and synchrotron features, if the multi-SNR cavity extends on both sides of the Plane, or no corresponding structure at all, in the case of a reheated SNR cavity limited to positive latitudes. Clearly, X-ray data lack a remarkable bright\linebreak X-ray area in the south and a bright spur. ROSAT~\cite{Snowden97} and Suzaku/MAXI~\cite{Tahara15} images reveal weak soft X-ray features, however they were judged too faint to be counterparts of the LNPS and overlooked. In the synchrotron domain, detailed WMAP and Planck measurements of diffuse polarized emission revealed a large series of arches and filaments, and among those the WMAP filament XII (see~\cite{Vidal15}) was proposed as a southern continuation of Loop I~\cite{Planck16}. According to the authors, the closed loop made from Loop I and filament XII could mark the tangential surface to the wide cavity blown by nearby SNRs (see Figure~\ref{fig:loops_cuts_new} and Section~\ref{eastwest} for discussion of such a~structure).

The situation has evolved very recently, with the first soft X-ray images recorded by the\linebreak e-Rosita instrument on-board the Spektr-RG observatory~\cite{Predehl20}. The e-Rosita images in the 0.3-2.3\linebreak(resp. 0.6-1.0) keV range shown in Figure~\ref{fig:presentatx} (resp. Figure~\ref{fig:figscheme}) clearly show a brightness enhancement in the south below the GC, and, in particular, an enhanced spur on the western side, quasi symmetric of the eastern NPS with respect to the GC (see next section). Brightnesses are significantly weaker than the north, though, especially at high negative latitudes, but the overall shape of the X-ray bright area corresponds well to an envelope of the southern FB, as shown in the top-right insert of Figure~\ref{fig:figscheme}, where $\leq10$~GeV gamma-ray are superimposed on the e-Rosita image (Figure taken from~\cite{Predehl20}. The very weak southern boundary at $\rm b \simeq-80$\fdeg can be seen in pale blue if one looks carefully at this composite image.

The southern X-ray envelope and spur, although rather convincing, are remarkably fainter, and the origin of the north-south asymmetry must be explained. Quantitative arguments were brought by~\cite{Sarkar19} who performed hydrodynamical simulations of the FBs and their interactions with the halo, and addressed the question of north-south X-ray brightness difference by varying halo characteristics. The author showed that the X-ray brightness difference may be explained by 20\% deficiency of halo gas density in the southern hemisphere in comparison with the northern one. According to the author, it would additionally explain why the southern FB is 5\% higher above the Plane than the northern FB: this would simply be due to the easier expansion in the less dense gas. Note that Sarkar's model addresses the case of FB generation due to quasi-continuous starburst at the GC and series of SNRs, while, today, the favored mechanism is accretion-fed explosive ejection. However, in both cases the expansion in the halo is strongly dependent on the halo density, and one may assume that his argument is also valid in the jet-like case. The result of~\cite{Sarkar19}, presented by the author as confirmation of a distant origin of the LPNS, was also used in an opposite way: arguing that there is no established reason for such a north-south difference of halo gas density, \cite{Panopoulou21} considered the result as evidence against the distant LNPS.

We performed a search in the recent literature for measurements related to the halo density and found some interesting results:

\pagebreak
\begin{enumerate}
\item \cite{French21} compiled and analyzed all past measurements of interstellar HI Lyman-series absorption with the Far Ultraviolet Spectroscopic Explorer (FUSE) and all HI Ly$\alpha$\linebreak absorption data towards extragalactic sources at latitudes $|b|\geq 20$\fdeg with the HST. The authors derived a larger fraction of intermediate velocity (IVC) gas in the northern halo. More quantitatively, they computed the sky covering fraction of IVCs as a function of their hydrogen column density and find that 70\% of sky covering fraction is reached for Log(N(HI)= 18.7~cm$^{-2}$ in the north and only Log(N(HI)\linebreak= 16.7~cm$^{-2}$ in the south. For the high velocity gas (HVC) the sky coverage is about the same, but there are more high column-density sightlines in the north.
\item \cite{Qu20} compiled and analyzed absorption by the SiIV ion towards AGNs using HST/COS data and developed a model of the halo warm gas associated with the ion. The authors found that the hot gas vertical scale height r$_{0}$ is larger in the north than in the south with r$^{N}_{0}= 6.3(+1.6,-1.5)$~kpc and r$^{S}_{0}= 3.6(+1.0,-0.9)$~kpc. If we use their results, this implies that between 1 and 10 kpc altitude, i.e. where the FBs and the heated X-ray gas volume are formed, the hot ionized halo gas density is significantly higher in the north (e.g., 25\% higher at 2kpc, and 80\% higher at 5~kpc).
\item \cite{Ashley20} analyzed HST-COS UV absorptions in the spectra of 5 quasars whose sight-lines cross the FBs with $30\geq|b|\geq 20$\fdeg\!. They found accelerating HI clouds in both hemispheres, but, interestingly, they also noticed anomalous blue-shifted clouds at b$\simeq 30$\fdeg and interpret these signatures as due to remnants of past outflows in front of the FBs.
\item It is finally possible to infer some asymmetry by taking a closer look at ROSAT ultra-soft X-rays images (0.25~keV). This easily absorbed emission is detected only towards the halo, and in the direction of the \emph{holes} in the local ISM. The emission reaches significantly stronger intensities in the north, even after exclusion of the LNPS area. In the past, this was interpreted as a sign for weaker absorption by cool gas at lower altitude, however, today measurements show that gas and dust columns in the northern hemisphere are similar or even higher than in the south, and this explanation in term of absorption does not hold. Instead, a stronger emission, must be preferred, i.e., the excess ultra-soft X-ray emission is potentially due to a higher hot gas density in the northern halo.
\end{enumerate}

This series of measurements and analyses all suggest the existence of non negligible differences between halo characteristics in the two hemispheres, and point to a larger density in the north, bringing support to Sarkar's explanation of the weak southern X-ray emission. Along with the X-ray brightness decrease, the model shows that radio features are also weaker in the south and have very different shapes. As a matter of fact, expansion in a low density medium implies lower energy transported by the shock wave and shock accelerated particles of less energy. If the southern halo is less dense, it may potentially explain the weakness of both southern X-ray and synchrotron emission. It remains that the rare, weak radio/micro-wave southern arcs are not well explained yet by a model. On the other hand, and interestingly, in an asymmetric configuration such like the one modeled by~\cite{Sarkar19}, the outer shock has reached a distance from the GC larger than in the north. In one of the model results, the distance reached in the south locates it slightly\linebreak beyond the Sun circle. In other words, the outer shock may have just crossed us in the south while the northern shock is still in front of us. This would mean that, at variance with the northern Loop I seen in the Galactic Center hemisphere at very high latitude, the synchrotron emission generated by the southern loop may appear differently. We will come back to this point in the discussion section.

\begin{figure}[!htbp]
\includegraphics[width=0.4\linewidth]{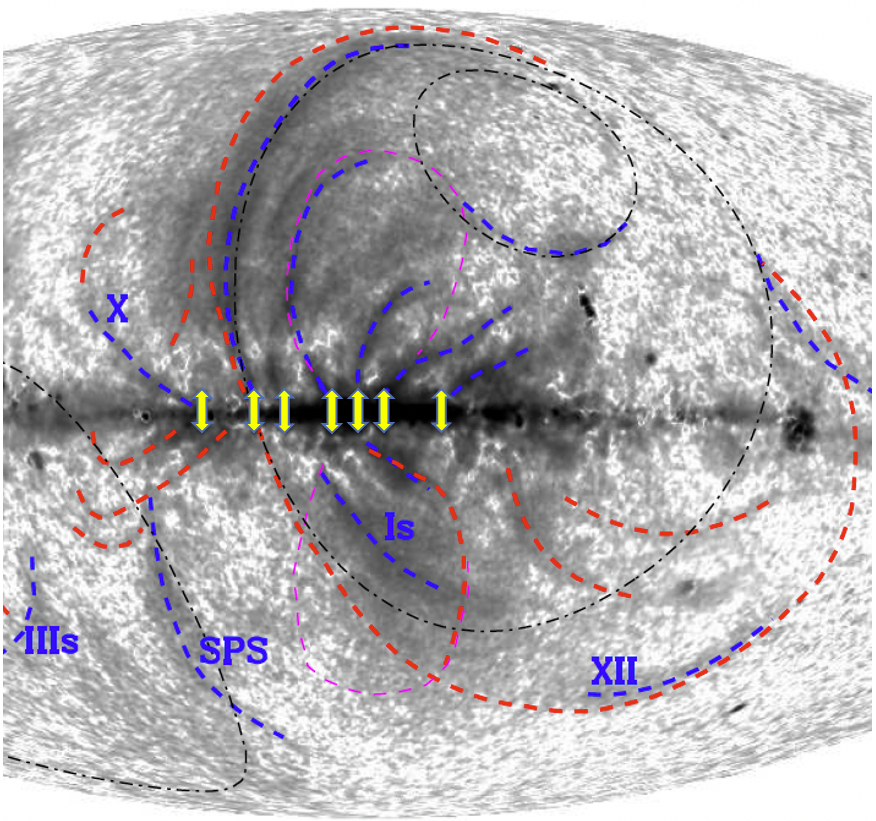}
\includegraphics[width=0.56\linewidth]{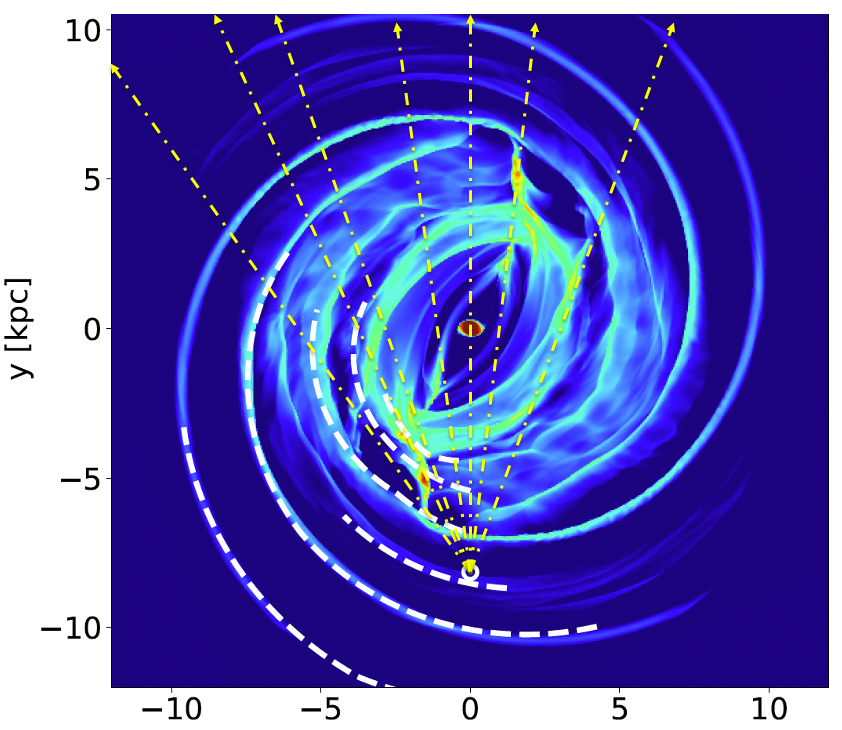}
\caption{\textbf{Left:} Fraction of the combined Planck/WMAP polarization intensity map (from~\cite{Planck16}). Superimposed as yellow arrows are the approximate bases (at latitude 0\fdeg\!\!) of the sharpest spurs/arches in the northern hemisphere. \textbf{Right:} High resolution hydrodynamical simulation of the Galactic gas adjusted to HI 21~cm data and the most recent Gaia and survey data (image from~\cite{Li21}). The Galactic plane is seen from the northern pole. Yellow dashed lines issued from the Sun have been superimposed on the image to indicate the longitudes of the low latitude ends of the spurs as defined in the left figure. One argument in favor of the GC origin of the LNPS is the coincidence between the direction of the low-latitude extremity of the LNPS ($\rm l\simeq25$\fdeg\!\!) and the direction of the high gas density area of the eastern 3-kpc arm (second direction from left). But there are also other interesting associations between spur and high gas concentration longitudes according to the new model of the entire Milky Way.}\label{fig:gasmodel}
\end{figure}

\section{The East-West asymmetry}\label{eastwest}

Another argument against a GC origin of the LNPS is the absence of east-west symmetry. Since the FBs are nearly symmetric around the GC, for homogeneous Milky Way disk and halo one would expect the X-ray and radio associated features to be nearly symmetric with respect to the meridian plane (zero longitude plane), as viewed from the Sun, which is not observed. In the north, X-ray data are characterized by a much stronger brightness in the east, with a fuzzier, weaker west counterpart. In the south, it is the opposite. Radio data do not show any high latitude ``spur'' in the west, and the low latitude arcs seen in polarized intensity (see Figure~\ref{fig:gasmodel}) are more numerous in the east and are all oriented in the same way.

Several recent works have shown strong evidence for an accretion-fed jet-type (collimated) event at the origin of the FBs, similar to jets in AGNs despite orders of magnitude of difference in power (e.g.~\cite{Ponti21, Cecil21}). The jets are collimated by the dense gas from the Central Molecular Zone (CMZ), which explains the FBs orientation at right angle from the Plane. It follows that, in case the LNPS is linked to the FBs, the observed east-west asymmetry seen in X-ray and radio must be related to posterior, differential evolution of the halo gas at large distances from the GC. Pursuing (in a different way) their early idea that the wide, 3~kpc radius very low density region of the inner disk is a remnant of GC past explosive events, \cite{Sofue21} recently suggested that the giant X-ray bubbles of shocked, expanding halo gas brighten where the flow encounters the 3~kpc molecular ring,\linebreak to form the X-ray then radio features. As a main argument, the authors use the fact that at the lowest latitude where the NPS can be detected, the longitude is $\rm l=22$\fdeg\!\!, i.e. very close to the direction of the tangent to the 3kpc ring.

Recent high resolution hydrodynamical simulations of the Galactic gas reproduce the 3~kpc ``ring'' in the gas distribution, as a consequence of the dynamical influence of the Galactic bar and the spiral arms. When based on a realistic barred Milky Way potential constrained by recent observations from Gaia and massive stellar surveys, the latest models from~\cite{Li21} reproduce most features in the observed HI 21~cm longitude-velocity diagram, including the Central Molecular Zone, the Near and Far 3-kpc arms, the Molecular Ring, and the spiral arm tangents. The gas distribution that comes out from these state-of-the-art models is strongly influenced by the bar and is roughly symmetric with respect to an axis oriented by about 25~degrees from the Sun-GC axis, i.e., it introduces a strong east-west asymmetry. One of the models is shown in Figure~\ref{fig:gasmodel}, and it can be seen that the gas over-densities, as viewed from the Sun vantage point, are seen at very different distances in the east and in the west. If, as suggested by Sofue and co-workers, the sources of the main soft X-rays brightness enhancements are associated with the gas flow encounter with 3~kpc structures, then it is clear that the ``east source'' is much closer (at $\simeq 5$~kpc from the Sun) than the ``west source'' at $\simeq 11$~kpc.

In an attempt to go one step further beyond the association between the LNPS base and tangential direction to the eastern ring, we have the same model of~\cite{Li21} and searched for potential associations between the longitudes of the polarized arches/spurs where they cross the Plane and the longitudes of the densest structures in the model (except the CMZ). In Figure~\ref{fig:gasmodel} are shown the approximate directions of the low latitude ends of the radio arcs/spurs seen in polarized emission and taken from~\cite[Figure~20]{Planck16}. We have reported this set of directions in the model map of~\cite{Li21}. There are several interesting associations that may mark the interaction of the shocked gas with disk gas concentrations, calling for further work in this direction. In any case, such an east-west asymmetry has important potential implications in the frame of the LPNS distant hypothesis: it will generate different characteristics of the propagation of the outer shock in the low altitude halo and modify the observed pattern as seen from the Sun vantage point, both in X-ray and radio, and create asymmetries.

\section{Microwave-radio and stellar polarization: the dual top-bottom structure hypo\-thesis}\label{topbot}

The LNPS synchrotron emission is highly polarized ($\simeq 40$-50\%), and spectacular images of the radio/microwave polarized fraction have been produced (see~\cite[Figure~20]{Planck16}, Figures~\ref{fig:gasmodel} and~\ref{fig:loops_cuts_new}). Information on the source distance based on the synchrotron signal has been searched for in several ways.
\begin{itemize}
\item A first estimate of the LPNS source distance is based on the amount of depolarization due to Faraday rotation in the foreground and the quantity of interstellar matter required to explain the depolarization (see~\cite{Wolleben06, Planck16, Dickinson18}). This technique has led to some constraints on the source location at low latitude (b $\leq30\fdeg$\!\!), where the radio signal is strongly depolarized, especially at low frequencies. A distance of several hundred pc is suggested by the most recent works (see~\cite{Dickinson18}. At higher latitudes there is no sign for depolarization, instead~\cite{Sun15} showed that the rotation measure (RM) of the NPS emission is negligible (see below other results from the same study). These studies suggest that there may be a dichotomy of the LNPS, with two different sources located at short distance (resp. higher distance) and viewed at high latitudes (resp. low latitude).

\item A second technique is the comparison between the polarization angle of the LNPS synchrotron emission and the direction of starlight polarization. A strong similarity between the starlight and LNPS polarization angles was well recognized for several decades in the case of nearby stars, but more recently the strong improvements in quality and sensitivity of stellar polarization measurements brought better constraints. \cite{Panopoulou21} used the latest optical data on polarization of stars closer than 1 kpc along with Planck and WMAP polarized synchrotron emission, also including the polarization angle of the dust thermal emission measured by Planck. The authors found that at $\rm b\leq30$\fdeg (their field~IA), there is no clear similarity in the directions of polarization, due to the presence of multiple structures. Above 30\fdeg\!\!, there is alignment between synchrotron and stellar polarization for stars located between 100 and 200~pc, and in two from three regions (fields IB and IC) the alignment holds at larger distances. In addition, where the dust is restricted to the first hundred parsecs, the polarization of the dust emission is aligned with the synchrotron. This is a very interesting result, and it strongly favors a very nearby source for the high latitude LNPS. According to their results, the authors suggest that the low and high latitude NPS emissions come from different sources. As a caveat, it must be said that the alignment is a necessary but not sufficient condition for a co-spatiality of the dust and the energetic electrons that emit synchrotron. If the magnetic field keeps the same orientation along large distances, both in the dusty region and beyond the dust, the synchrotron emission may originate from beyond the dust, while the angle between the polarization directions of the synchrotron emission and the stellar light remains the same. This is not precluded in the two regions mentioned above (fields IB and IC), since the good alignment holds until the limit distance for the alignment study, about 600~pc. The conclusion drawn, namely that above 30\fdeg latitude the synchrotron emission is generated as close as $\simeq100$~pc, may not be valid in those fields. It remains valid in the highest latitude field ID.

\item A third and very promising technique is Faraday tomography. \cite{Sun15} used the first data from the GMIMS Survey using 2048 frequency channels from 1280 to 1750~MHz to obtain a Faraday Rotation Measure (RM) of the NPS and in adjacent regions. They combined the RMs with Faraday depth maps of the entire Galaxy based on extra-galactic sources assembled by~\cite{Oppermann15}. From the comparison between the two measurements and using the difference between data directed towards the NPS and data in adjacent regions, they estimated the Faraday depth in front and beyond the NPS and used those estimates to draw several conclusions: -below $\rm{b}=44$\fdeg the emission is very likely distant, or the large scale magnetic field must have a reversal;-above this latitude, the Faraday depth is close to zero and the LNPS emission must be local and originate within several hundreds parsecs. Such a pioneering study is quite interesting; however, at high latitudes the measurements and differences are close to the uncertainty level, and there are several simplifying assumptions, such like the assumed similarity between all physical quantities in adjacent regions outside and inside the NPS, an assumption that may not be valid if the NPS corresponds to the periphery of shocked expanding hot gas, and the additional assumption of perfect mixing between thermal and non thermal emitting gas within the~NPS.
\end{itemize}

As a conclusion, interestingly all three recent studies based on polarization point to a double source, a low-latitude and distant source and a high-latitude and nearby source. Again, we must note that at high latitudes the total column of interstellar dust is reached at very short distance (see the dust map section below), making the argument of negligible depolarization or Faraday rotation as a sign of proximity of the emission disputable. We must also note that such a dual source is in contradiction with the assumption that there is a unique giant reheated cavity bounded by Loop~I and the continuation of Loop~I under the form of Loop~XII (from~\cite{Vidal15}) (see Section~\ref{Reheated} and Figure~\ref{fig:loops_cuts_new}).

A dual structure is contradicted by X-ray data. At variance with polarization measurements, X-ray data favor the continuity of the LNPS source properties from low to high latitudes, because there is no discontinuity nor sign of inhomogeneity of the recorded spectra. \cite{Akita18} analyzed the entirety of Suzaku spectra in the direction of the LNPS and found that all spectra can be fitted with a $\simeq 0.3$~keV gas, with no need for any discontinuity, nor even variation except for the emission measure. Such homogeneity is in agreement with the previous analyses based on Suzaku, XMM-Newton and Swift observations in various directions within both the LNPS and the northern FB contour (\!\!\cite{Kataoka13,Kataoka15,Tahara15}). More recently, and totally independently, the analysis of the spectra recorded with the nano-satellite HaloSat led to the same conclusion. HaloSat (\!\!\cite{Kaaret19}) was dedicated to the diffuse emission from the hot gas in the halo, but in parallel did a series of observations along the bright LNPS from galactic latitudes $+15$ to $+74$\fdeg and found a total continuity in the spectral shape~(\!\!\cite{LaRocca20}).

\begin{figure}[tbp]
\includegraphics[width=0.8\linewidth]{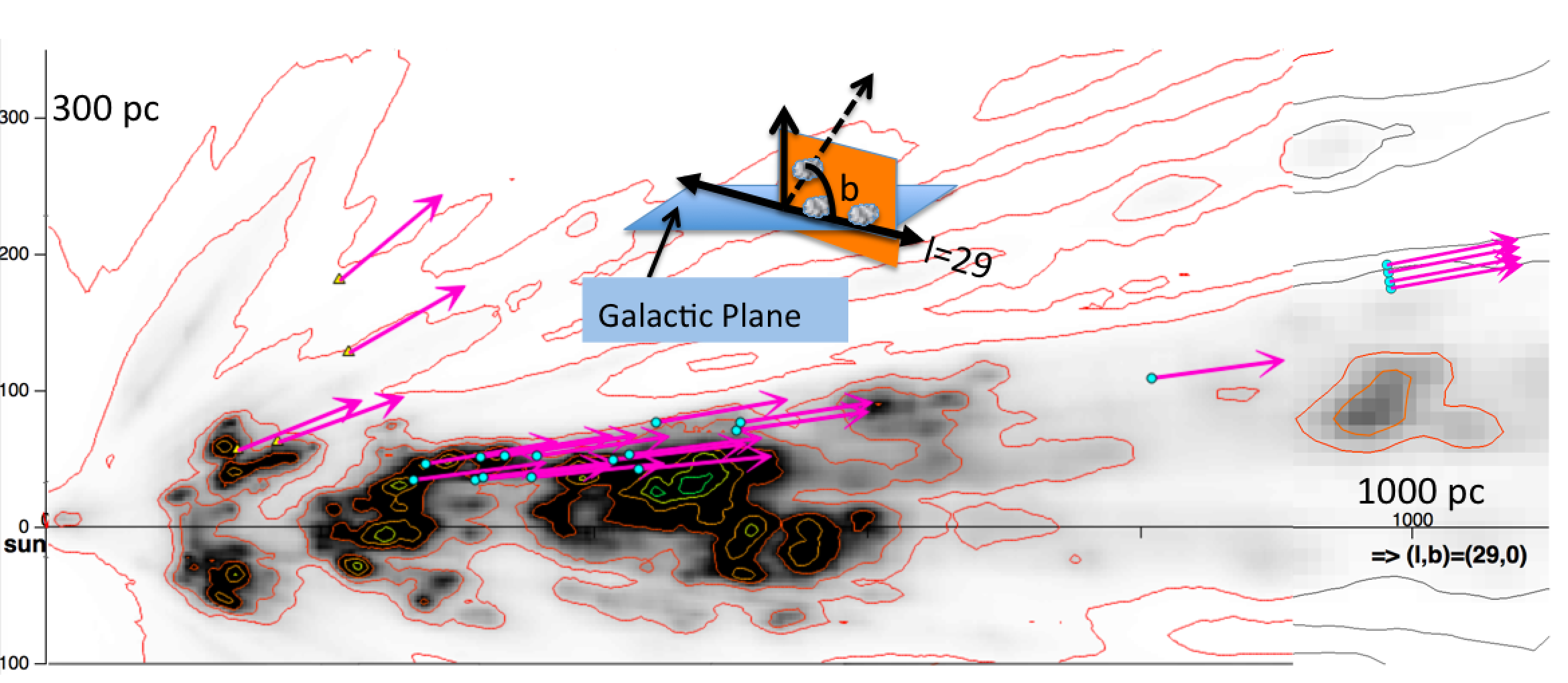}
\caption{Distances at which measured X-ray absorbing columns are reached towards the NPS. The figure is the image of the dust extinction density distribution in a vertical plane containing the Sun and the direction of the northern Galactic pole, and oriented along the Galactic longitude $\rm l=29$\fdeg (the plane colored in orange in the small inserted scheme). The blue circles are associated with a series of 19 XMM-Newton spectra recorded in this plane at galactic latitudes comprised between $+5.6$ and $+11.2$\fdeg\!\!~\cite{Lall16}. The spectral analysis provided for each direction the value of the column of gas absorbing the NPS signal. After conversion to extinction, the map is used to calculate the distance along the corresponding line-of-sight at which this value is reached, i.e. the minimum distance to the region responsible for the NPS emission. For each pointing the corresponding location is indicated by an arrow whose location marks this limit and the region beyond which the emission starts to be produced. Note that in the case of the four circles at right in the figure the computed distance would locate them out of the figure. The four yellow triangles correspond to other additional measurements towards the NPS, however the longitudes of the probed directions are not $+29$\fdeg (see text).}\label{fig:absorbX}
\end{figure}

\section{Distances from X-ray data}\label{distfromxray}

\subsection{Using global spectral shapes}

More quantitative estimates of the distances to the LNPS source regions can be obtained from spectral information on the absorbing foreground. Several X-ray spectra have been recorded towards NPS regions, and various conclusions have been drawn depending on the probed region and the spectral analysis. XMM-Newton then later on Suzaku were used to record spectra towards the NPS above $\rm b=+20$\fdeg (\!\!\cite{Willingale03,Miller08}). The data analysis suggested a significant absorbing column on the order of the estimated full Galactic column. In the frame of a wider study of the whole X-ray emission supposedly associated with the Fermi bubbles, \cite{Kataoka13} recorded and analyzed the Suzaku spectra for 8 sightlines towards the high latitude NPS ($43\fdeg\!\!\leq{\rm b} \leq48\fdeg$\!). Using a single component for the NPS and Galactic halo (in addition to the Local Bubble) the authors found that the column of the cold absorbing foreground is on the order of the full Galactic column. This was confirmed by later analyses of the same data augmented by several additional Suzaku and SWIFT spectra (\!\!\cite{Kataoka15,Tahara15}).

\begin{figure}[!hbp]
\includegraphics[width=0.90\linewidth]{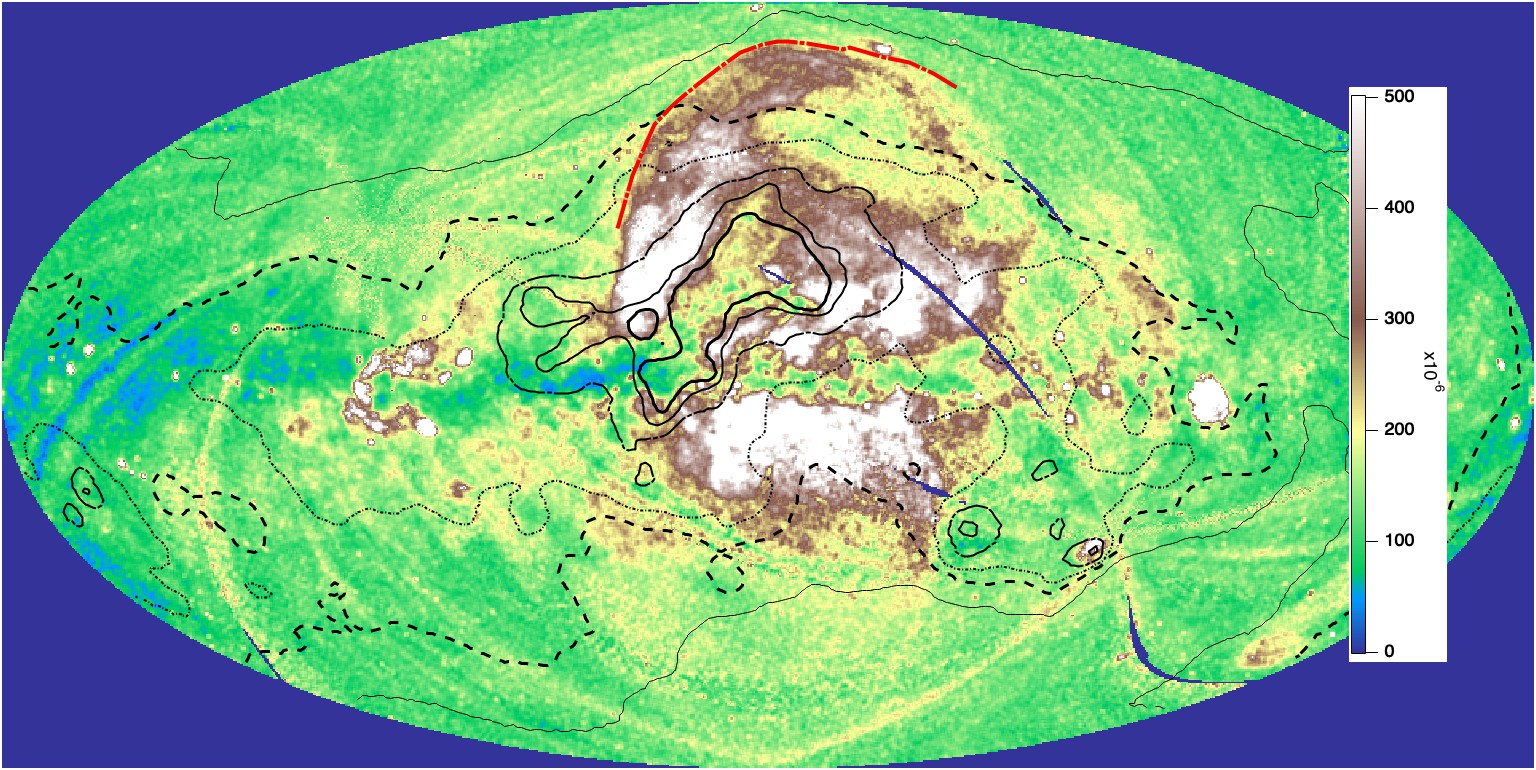}
\includegraphics[width=0.90\linewidth]{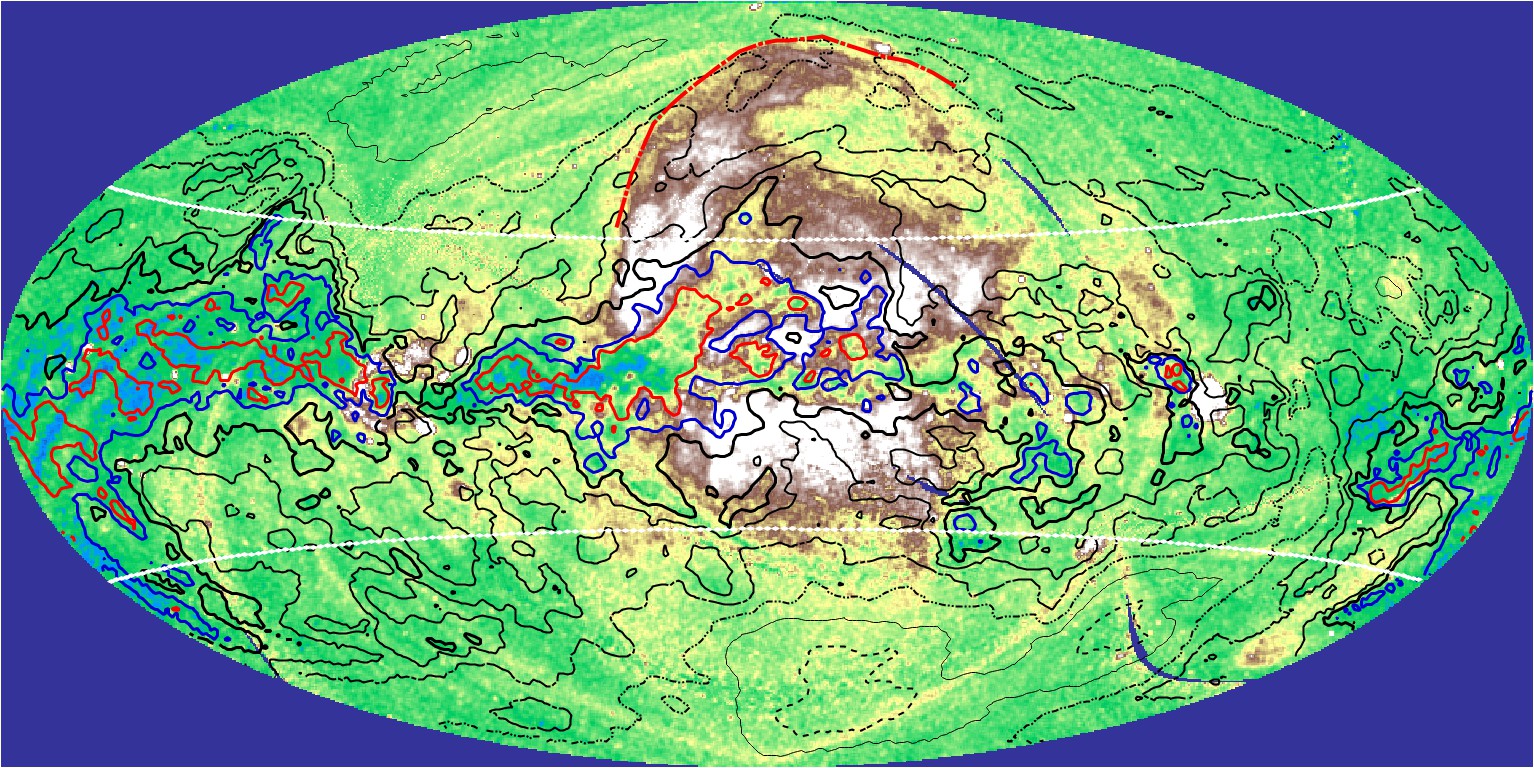}
\caption{\textbf{Top:} Isocontours of dust extinction in the visible, integrated between 0 and 100~pc (\!\!\cite{Vergely22}), superimposed on the ROSAT 0.75~keV map (\!\!\cite{Snowden95}). Contours are drawn for A$_{V}=0.02, 0.03, 0.04, 0.06, 0.08$ and 0.1 mag. The LNPS eastern limit is shown in red. \textbf{Bottom:} same as top, but the extinction is integrated between 0 and~800 pc. and contours for A$_{V} =0.04, 0.06, 0.1, 0.2, 0.4, 0.8$ (blue) and 1.5 (red) mag. White lines mark the latitudes above which the integration is limited by the boundaries of the 3D map and does not reach 800~pc (the distance reached is between 400 and 800~pc). At those latitudes, however, beyond 400~pc the extinction is negligible. Note the detailed anti-correlation with the X-ray brightness for most contours in the bottom map, at variance with the 100~pc integration at top where only a very weak anti-correlation is seen for the Aquila Rift at low latitude.}\label{rosat75dust}
\end{figure}

\begin{figure}[!htbp]
\includegraphics[width=0.90\linewidth]{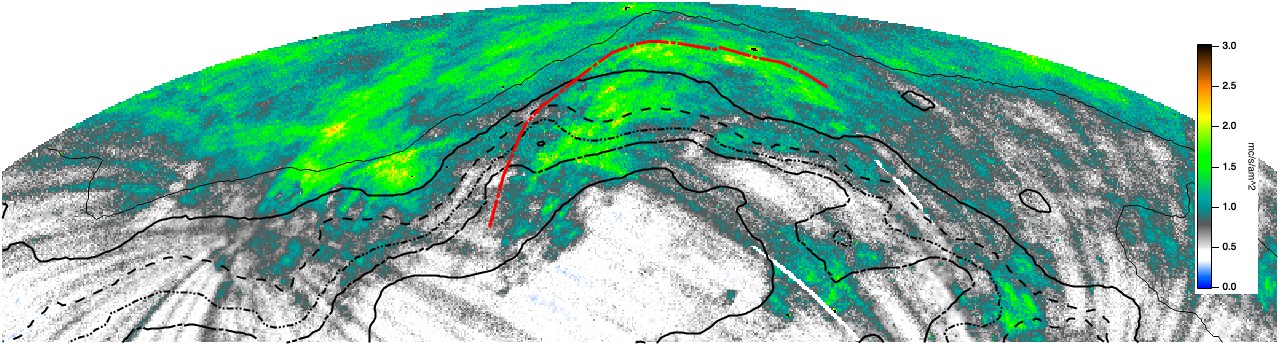}
\includegraphics[width=0.90\linewidth]{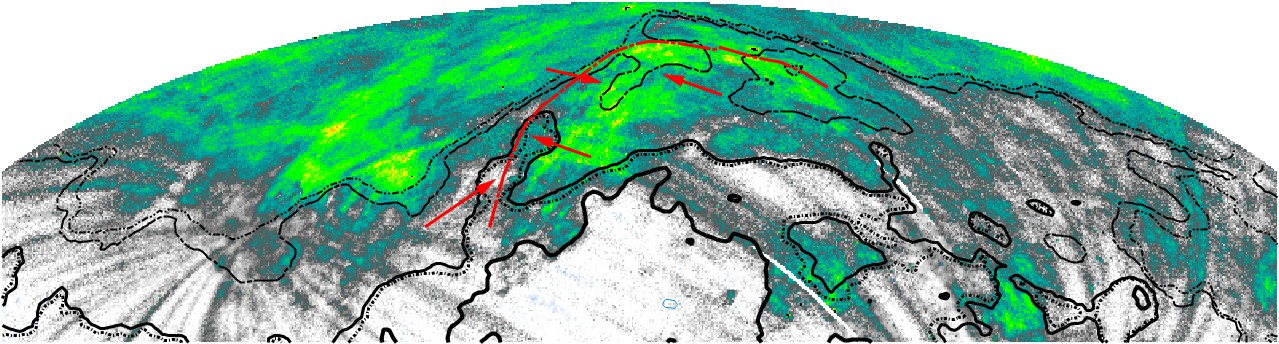}
\caption{\textbf{Top:} Isocontours of dust extinction in the visible, integrated between 0 and 100~pc (\!\!\cite{Vergely22}), superimposed on the ROSAT 0.25~keV map (\!\!\cite{Snowden95}). Note the strong change of the color scale. The recorded signal is restricted to high latitudes directions for which only weak absorption occurs, i.e. from the halo and the top part of the LNPS. The LNPS eastern limit is shown in red. It allows to distinguish the contribution to the signal of the LNPS from the one of the halo. Contours are drawn for A$_{V}= 0.02,$ 0.025, 0.03, 0.035, 0.04, 0.06~mag.\linebreak\textbf{Bottom:} same as top, but the extinction is integrated between 0 and 200 pc and contours at $\rm A_{V}= 0.035,$ 0.04, 0.09, 0.09, 0.2~mag. Note the appearance of anti-correlation for several contours in the bottom map. Also note the absorption due to the elongated cloud overlapping the LNPS ridge (e.g. at locations indicated by red arrows).}\label{rosat25dust}
\end{figure}

An interesting region is the low latitude part of the X-ray NPS around $\rm b=+10$\fdeg\!\!, a region characterized by an abrupt disappearance of the signal below $\simeq 6$\fdeg latitude (see Figure~\ref{fig:presentatx}. A series of XMM-Newton spectra were recorded with the goal of determining whether this disappearance is due to increased interstellar matter absorbing column (absorption-bounded case) or to the true lack of emission (emission-bounded case) below this latitude (see~\cite{Lall16}). The analysis of the spectral features and intensities showed that the absorption-bounded case is the actual situation, and a series of absorbing columns were measured between $+5.6$ and $+11.2$\fdeg\!\!, all around the $l=+29$\fdeg longitude. The model used for this low-latitude part of the NPS was a composite model of the NPS itself, the higher energy emission of the \emph{bulge} and the pre-determined small contribution of the Local Bubble at the softer energies, and the absorption was assumed to be entirely due to cold gas. Absorbing columns on the order of 5 to 50 10$^{20}$~cm$^{-2}$ were obtained from the spectral fits. After conversion of gas columns to dust extinctions and comparisons with existing 3D dust maps, estimates of the distance from the Sun at which the measured absorbing column is reached could be obtained for each sightline and a first estimate of the minimum distance to the X-ray source was 300~pc; however the good correlation of the absorbing columns with the optical thickness of the intervening dust deduced from Planck, generated from the Sun to infinity, was suggesting a larger distance. More recently, \cite{Das20} used the same measured absorbing columns and updated 3D extinction radial profiles based on PanSTARRS and 2MASS to perform a new conversion from gas to dust and estimate distances at which the absorption takes place. In order to visualize the absorbing clouds, we interpolated in their table~1 to compute the distances at which 100\% of the converted gas column is reached. In Figure~\ref{fig:absorbX} are shown the locations of such limits, superimposed on the image of the extinction density distribution in a vertical plane containing the Sun and oriented towards $\rm l=29$\fdeg\!\!. The extinction distribution, here shown up to 1050~pc, is deduced from a recent inversion of individual measurements of individual extinctions from GAIA eDR3 and 2MASS photometric data analyses and from ground-based spectroscopic surveys~\cite{Vergely22}. Since the 19 XMM-Newton lines of sight are within $0.7$\fdeg from the represented plane, the image gives a close representation of the dust along all of them. Four limit distances are higher than 1 kpc and fall out of the map and are shown at the map boundary. 4 additional points at high latitude are also shown, from two different works and are discussed below.

Two comments can be made from such the X-ray and dust data: first, the limits are either close to locations beyond which the quantity of dust is very low or beyond all the dust. Given the large uncertainties on the gas to dust conversion factor and on the absorbing columns (see below), this pattern is compatible with a source of the emission from a volume beyond or above the mapped dust clouds (this would happen in case the absorption is slightly underestimated); second, and more important, even if the absorbing columns (and minimum distance) were slightly overestimated, the minimum distance to the emission source seems to increase with latitude among the 19 low-latitude sightlines, a pattern certainly not suggestive of a source region located close to the Sun, and certainly not at distances on the order of 150~pc above the Sco-Cen associations. Instead, the pattern suggests a source located above the dust layers or tangential to the long series of dust clouds from the Sun to about 1000~pc. In Figure~\ref{fig:absorbX} we have added four limit distances based on measurements of X-ray absorbing columns from~\cite{Willingale03,Miller08},\linebreak and a conversion factor of 5 10$^{21}$~cm$^{-2}$ per E(B-V) mag. The longitudes for these measurements are significantly different from $\rm l=29$\fdeg\!\!, however dust images in the corresponding planes show a close pattern. The location of the limits gives an idea of the constraints, or more precisely, of the absence of constraints on the distance to the emission: as a matter of fact the absorbing column places the limits beyond the reconstructed dust, implying that the source may be everywhere at larger~distances.

\subsection{Using anti-correlation between soft X-rays and dust opacities}

The inter-comparison of diffuse soft X-ray images in various energy bands provides primarily information on the hot gas temperature and density, but may additionally provide constraints on the distance to the source, provided one has some preliminary information on how the absorbing gas is distributed along the line of sight. This is due to the enormous variation of the absorption cross-section per H atom $\sigma_{X}$ over the soft X-ray energy range. In the case of the 0.1 to 2 keV ROSAT maps (\!\!\cite{Snowden95}), $\sigma_{X}$ decreases by a factor above 500 with increasing photon energy. This is why 0.25~keV data reveal solely the \emph{unabsorbed} or very weakly absorbed emission from hot gas in the local bubble and from the high latitude halo (N$_{H}$ columns on the order of\linebreak 10$^{19-20}$~cm$^{-2}$), while 0.75~keV can be used to detect hot gas beyond more opaque regions (columns on the order of up to 10$^{22}$~cm$^{-2}$). The observed LNPS soft X-ray spectrum becomes harder at low latitude (until signal disappearance below $\simeq$ 6\fdeg\!\!), while at high latitude the Galactic ISM columns become small enough to allow detection in the softer range. This explains the color variation in the e-Rosita map of Figure~\ref{fig:presentatx}. Thanks to Gaia and ground-based surveys, the increased availability of 3D information on the Galactic ISM, and in particular on dust extinction allows to study in more detail the distance to the emitting gas. Assuming a uniform dust to gas conversion factor, it is possible to build maps of the columns of absorbing gas integrated from the Sun to a given distance, and to compare the resulting images with diffuse X-ray background images. Against a homogeneous background source, variations of the absorbing columns should be anti-correlated with the observed, absorbed signal. We have taken advantage of the new 3D extinction maps based on Gaia distances and photometry to do this exercise (\!\!\cite{Vergely22}), and Figures~\ref{rosat75dust} and~\ref{rosat25dust} illustrate some of the results. An entire analysis is beyond the scope of this article.

Figure~\ref{rosat75dust} shows two examples of comparisons between the 0.75~keV diffuse emission and dust extinction in the visible A$_{V}$. The extinction between the Sun and 100~pc (resp. 800~pc) is shown under the form of iso-extinction contours that can be compared with the X-ray intensity pattern. While no or very little anti-correlation can be seen for the short distance extinction, the 800 pc extinction map is clearly revealing detailed anti-correlation patterns. This confirms the previous analysis of XMM data and extends to all longitudes the conclusion reached at $\rm l=+29$\fdeg\!\!, namely that the emission is generated well beyond the Aquila Rift and very likely beyond 800 pc. Figure~\ref{rosat25dust} is of similar nature for the ultra-soft energy map (0.25~keV). In this range, X-ray emission is detected only at high latitude where absorption is very weak, and the LNPS ridge is no longer a sharp limit as at higher energy. We show the extinction iso-contours for 100 and 200 pc only. Again, no anti-correlation is visible at 100 pc, showing that the absorbing medium is located at further distance. Conversely, between 100 and 200~pc anti-correlations appear, in particular the absorption effect of an elongated cloud covering the LNPS ridge is visible. This cloud, whose HI counterpart has been already discussed by~\cite{Sun15}, is also visible as a weak dark feature in Figure~\ref{fig:absorbX} at $(\rm l,b)=(+29\fdeg\!\!, +60\fdeg\!\!)$. Isocontours computed for longer distances are similar to those for 200~pc, because there is no more dust beyond 200~pc at such high latitudes. This pattern suggests that at high latitude the X-ray signal is generated beyond the totality of the dust.

\subsection{Using line ratios}
After the published analysis of the early LNPS soft X-ray spectra recorded with XMM-Newton and Suzaku above $\rm b=+20$\fdeg (\!\!\cite{Willingale03,Miller08}), some peculiar features in the spectra attracted the attention. \cite{Lall09} noticed a Doppler shift of the OVII and NeIX He-like triplets (made from unresolved resonance, intermediate and forbidden lines) and interpreted it as potential forbidden line enhancement due to charge-exchange (CX) cascade emission, a mechanism potentially at work at the interface between the NPS hot gas and the cold matter of the disk. Later on, and almost at the same time as the XMM-Newton study quoted above, \cite{Gu16} re-reduced and analyzed in detail all previous XMM and Suzaku data above $\rm b=+20$\fdeg\!\!, introducing in the spectral analysis a potential absorption by ionized gas, in addition to the absorption by dense, neutral matter. Compared to the spectral modeling of~\cite{Lall16}, the analysis was made differently. \cite{Gu16} used data recorded at the same latitude but out of the NPS to perform a preliminary determination of the emission from the LB and the one from the Galactic halo (at such latitudes the halo emission is a non-negligible contributor, which not was not the case for the NPS between 6 and 11 degrees latitude where the emission from the bulge is the main additional contributor). The authors built model spectra for various combinations of thermal+CX emission and absorbing columns of both neutral and ionized interstellar gas and found that a high absorbing column of hot (0.17-0.20~keV) ionized gas is necessary to reproduce simultaneously the enhanced OVII and Ne IX forbidden-to-resonance ratios and high OVIII Ly$\beta$ line relative to other Lyman series. They found a minor CX contribution to the emission on the order of 10\% only. The similarity between fitted model spectra and data obtained by the authors in this case is very strong. They concluded that the high temperature and the high columns of this absorbing ionized gas (on the order of 3-5 10$^{19}$~cm$^{-2}$) play in favor of halo gas absorption over large distances (5 to 8~kpc). If the hot gas column were distributed over distances $\leq 1$~kpc, the size of the proposed nearby cavities (see also below), its average density would be higher than 0.01~cm$^{-3}$. This is above average hot gas densities in hot super-bubbles according to hydrodynamical models (see e.g.~\cite{MacLow89}), instead long (above 5-6~kpc) paths through hot halo gas are more likely. The conclusion of this overlooked study is in agreement with the conclusion deduced from the NPS low latitude edge, however, in the case of hot ionized gas absorption, the favored distances reach Galactic scale sizes.

\section{A re-heated superbubble within the Local Arm?}\label{Reheated}

\begin{figure}[!htbp]
\includegraphics[width=0.90\linewidth]{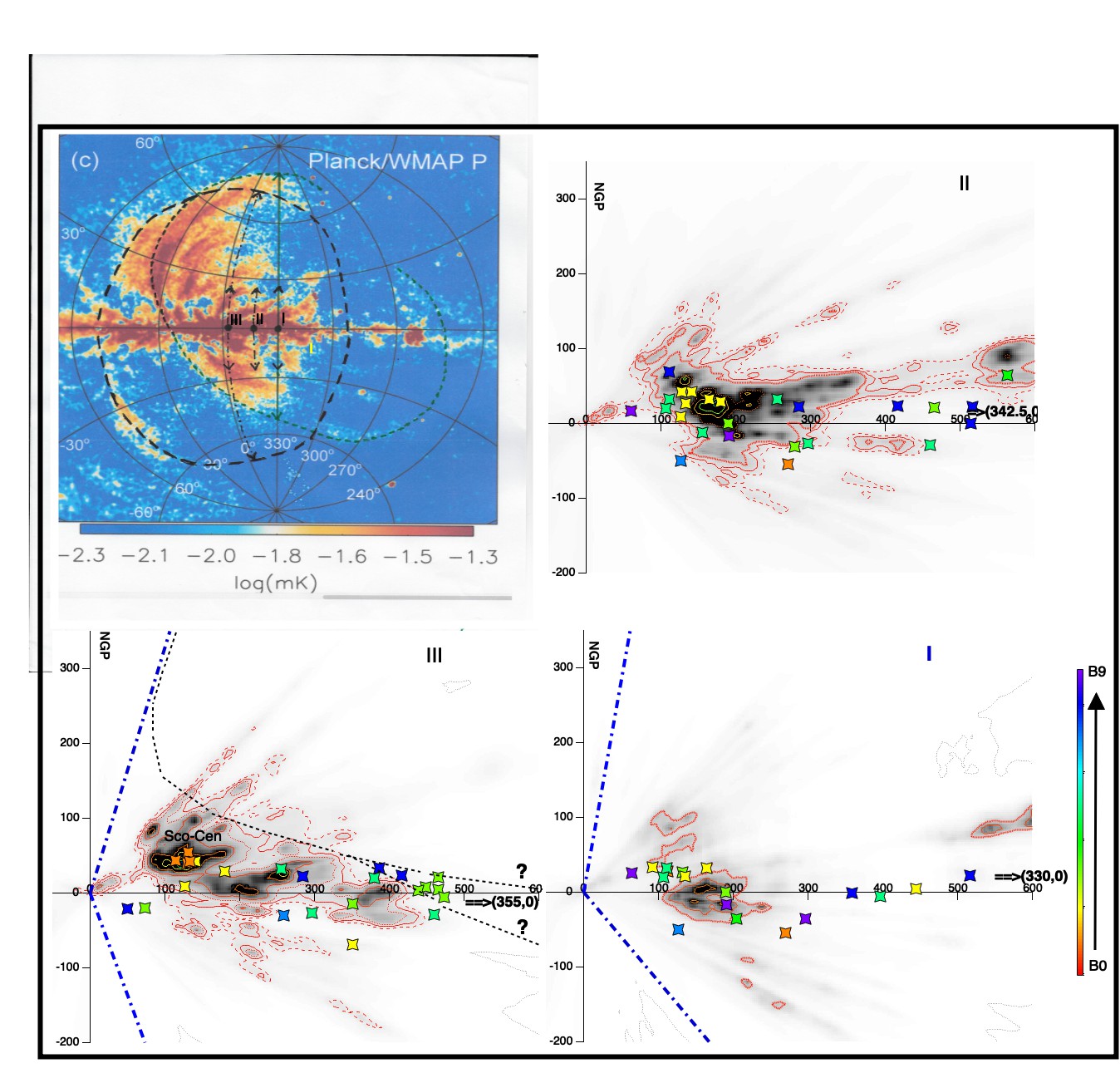}

\caption{\textbf{Top left:} combined Planck/WMAP polarization intensity, showing the LNPS and all other arches/spurs (\!\!\cite[Figure 21(c) from]{Planck16}). The contours of the local shell proposed by~\cite{Planck16} are shown by a thin dashed black line. We have superimposed as a thick dashed black contour the local shell due to Loop I interaction with the Local cavity proposed by~\cite{Wolleben07}. We have selected three meridian lines, marked by I to III and black arrows. I corresponds to the center of the~\cite{Planck16} shell, III to the center of the shell proposed by~\cite{Wolleben07}, and II is the intermediate longitude. \textbf{Right:} dust extinction density images in the three vertical meridian planes containing the Sun, the north pole direction and oriented towards Galactic longitudes 330\fdeg (I), 342.5\fdeg (II) and 355\fdeg (III), i.e. those containing the directions marked by black arrows in the left figure. Massive B stars from the ALMA catalog (\!\!\cite{Pantaleoni21}) and located within 10 degrees from each plane are shown in projection onto the plane (note that some targets appear on two figures). Colors refer to the stellar type, as indicated on the color scale at bottom right. Blue dot-dashed lines indicate the tangential directions to the boundaries of the shells I and III. A potential solution for the near side boundary of a giant reheated cavity of hot gas is indicated by dashed contours in III (see text).}
\label{fig:loops_cuts_new}
\end{figure}

Figure~\ref{fig:loops_cuts_new} shows the diffuse polarized emission from combined Planck and WMAP data (taken from~\cite{Planck16}). There are several loops and arches whose distances have been discussed by the authors, who also proposed a prolongation of Loop I in the south and in the west to form the closed loop indicated by a white contour and marking the synchrotron-emitting tangential surface to a giant cavity of recently reheated hot gas, centered at several hundred parsec. As a matter of fact, the existence of strong depolarization below $\rm b=30$\fdeg (\!\!\cite{Wolleben06, Planck16, Dickinson18}), the clear imprints of Aquila rift absorption in ROSAT maps and first evidences for the absence of a very nearby huge cavity in 3D maps of the ISM (\!\!\cite{Puspitarini14}) led~\cite{Planck16} to propose a two-steps scenario to explain both diffuse X-rays and synchrotron radiation, namely (i) the creation more than 10 Myrs ago of a huge hot gas cavity blown by a large series of SNRs, and centered at several hundred parsecs, and (ii) the more recent (less than 2~Myrs ago) reheating of this cavity by a second SNR episode. The giant shell boundary crosses the Plane at $\rm l\simeq+25$\fdeg\!\!. In such a configuration the blown and reheated hot gas cavity encircled by the shell is clearly asymmetric. We have superimposed on the figure is a second close loop proposed and modeled by~\cite{Wolleben07}. According to this work, this shell marks the interaction between Loop I and the Local Bubble. As part of the figure are drawn images of the dust distribution in vertical (meridian) planes along the longitudes of the shell centers (marked by I and III respectively). The directions of the high and low latitudes tangential directions to the respective shells are also indicated. The proposed cavities should be bordered by these two directions and centered at several hundred parsec in the first case, and much closer in the second case. According to the previous section, the strongest X-ray-based argument against such a double episode of large cavity creation and reheating, as proposed by~\cite{Planck16}, is the requirement for a very large column of hot ionized gas in front of the LNPS, deduced from specific line intensities by~\cite{Gu16}. As discussed above, this is hardly compatible with a cavity size of about 1~kpc or less. Assuming that such an inference is not granted, because it is is based on a single type of spectral analysis, we consider the re-heating scenario in the light of the above more numerous constraints on the distance to the X-ray source.

First of all, we note that this is not an unusual configuration, as series of SNRs occur episodically and super-bubbles are observed here and there. Moreover, the recent detection at Earth of the $^{60}$Fe isotope (whose lifetime is on the order of 2 Myrs) is reinforcing the existence of such a scenario because it is naturally explained by one or more very recent SNR explosions at very short distance from the Sun, as argued by~\cite{Breitschwerdt16}. The authors have traced back the Sco-Cen stellar association and modeled the evolution and impact of its members. Interestingly, they have shown that two of the most recent supernovae from the group have very likely exploded 1.5-2~Myrs ago within 150~pc and may be at the origin of $^{60}$Fe, while the older members may have produced 14-20~supernovae and have gradually blown the Local Bubble. This is similar to the re-heating scenario of~\cite{Planck16}, replacing the Local Bubble by a more distant cavity centered at longitudes between $\rm l=330$\fdeg and $\rm l=0$\fdeg. In the case of the LNPS, a super-bubble centered at several hundred parsecs (very likely more than 700~pc as we discussed in Section~\ref{distfromxray}), and culminating above the Sun at b=+80\fdeg has a minimum radius on the order of 600~pc and a volume at least one order of magnitude above the one of the LB. Models of super-bubbles blown by stellar winds and supernovae predict that maintaining such gigantic hot gas cavities requires a frequency of massive star deaths on the order of five per Myr during more than 5~Myrs (see, e.g.~\cite{MacLow89}). In this case, OB associations having given rise to the bubble must have left groups of early and late B stars, and there should be a huge cavity in the ISM (see, e.g. evolutive MHD models of~\cite{DeAvillez05}, well suited for the Sun vicinity). Both types of signatures are difficult to identify, however thanks to Gaia there has been some progress in the two cases.

First, an extensive new catalog of massive stars within 3kpc has been established and very recently presented by~\cite{Pantaleoni21}. Importantly, for most objects from the catalog located within $\simeq 1$~kpc rather accurate Gaia parallaxes are available. We have updated the Gaia DR2 distances given in the catalog to their eDR3 re-evaluations and extracted all massive O, B stars from the catalog with longitudes between $\rm l=320$\fdeg and $+5$\fdeg\!\!. The objects are superimposed on the dust images of Figure~\ref{fig:loops_cuts_new}. There are two groups of associations, the nearby, well-known Sco-Cen association and a second group of older stars at about 450~pc close to the Plane. Most stars from the former group are still embedded within their massive parent molecular clouds. There are relatively small\linebreak(20-60~pc wide) cavities associated with the youngest and most massive objects, in particular the cavity associated with $\rho$Oph (\!\!\cite{Robitaille18}). The 3D dust map resolution is too poor to reveal the details of the smaller cavities, only a few are visible. These young objects may have been a potential source of reheating of a more ancient super-bubble. The second group of stars at about 450~pc contains about ten B stars which could belong to the group having first produced the giant cavity. The number of objects, however, appears rather low in comparison of expectations for a huge cavity. From the geometrical point of view, the huge reheated cavity could be bounded in the way schematized in Figure~\ref{fig:loops_cuts_new} for the $\rm l=355\fdeg$ vertical plane. In this plane one can see that this cavity could extend in the southern hemisphere at large distance (see the question marks in Figure~\ref{fig:loops_cuts_new}), following the polarization full loop proposed by~\cite{Planck16}. However, cloud distributions at other longitudes, e.g. at $\rm l=342.5$\fdeg in this figure, or at $\rm l\simeq+29$\fdeg in Figure~\ref{fig:absorbX}, i.e. close to the longitude at which the outer boundary of the cavity crosses the Plane, are more difficult to reconcile with a southern extension to form a giant north-south cavity.

\begin{figure}[!htbp]
\includegraphics[width=0.5\linewidth]{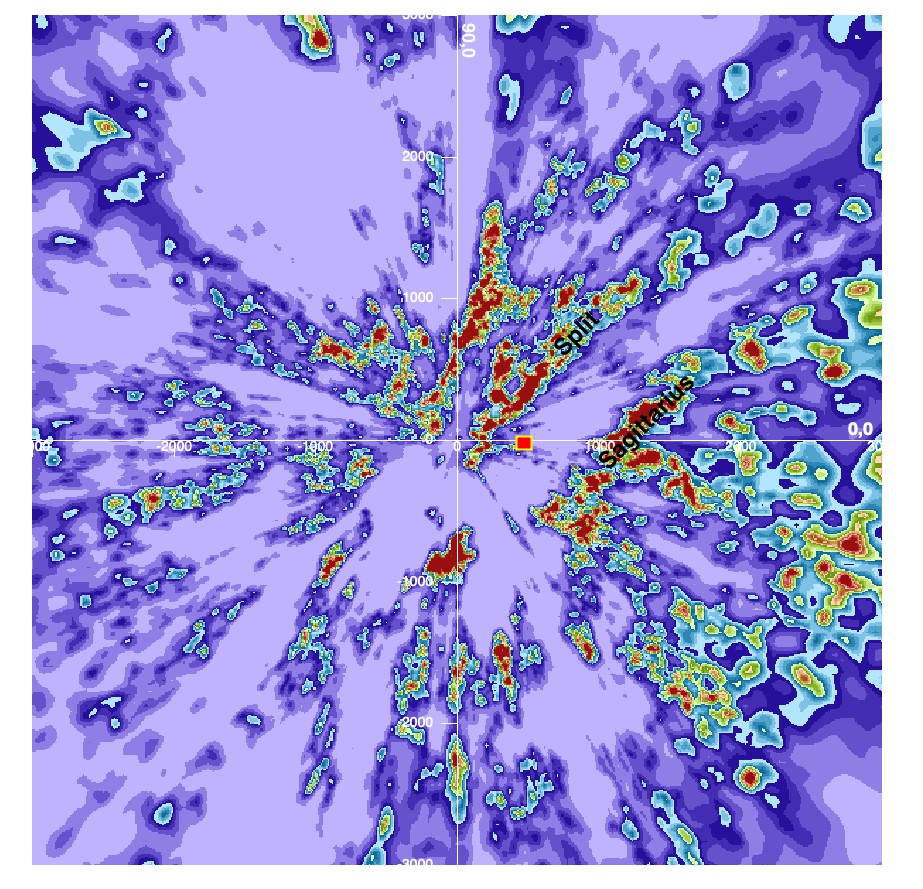}
\caption{Dust distribution along the midplane within 3~kpc from the Sun. The Galactic center direction is to the right, units are pc. The location of the group of massive B stars from the ALMA catalog (\cite{Pantaleoni21}) considered as a potential source of a nearby superbubble is indicated by a red square.}\label{galplane}
\vspace{-10pt}
\end{figure}

On the second aspect of ISM distribution, we show in Figure~\ref{galplane} the distribution of the dust within 3~kpc from the Sun as deduced from the very recent mapping we already mentioned (\!\!\cite{Vergely22}). The \emph{old} B star cluster is located close to the boundary of the giant dust-free region separating the Local Arm (more precisely its inner part, the conspicuous \emph{split} (\!\!\cite{Lall19}) from the $\geq 1$ kpc distant Sagittarius--Carina outer part. There is no indication of any large scale cavity or even imprinted large scale curvature in the dust distributions around the star group, instead the chains of dust clouds follow the same orientation everywhere. As a conclusion, within the limits of current knowledge, there are no hints of any kind of a super-bubble creation within the Local Arm close to the Galactic Center direction, although all arguments are not very strong. We do not discuss further a super-bubble generated within Sagittarius (at $\simeq 1.2$~kpc), whose radius would be above 1.1~kpc. Although this can not be excluded, we now reach a probability on the same order as the interaction of a shock launched by FB-type events with the disk material at and beyond the 3~kpc ring (potentially at 3-4~kpc from the Sun, see Figure~\ref{fig:gasmodel}).

\section{Summary and discussion}\label{sect7}

\begin{figure}[!hb]
\includegraphics[width=0.48\linewidth]{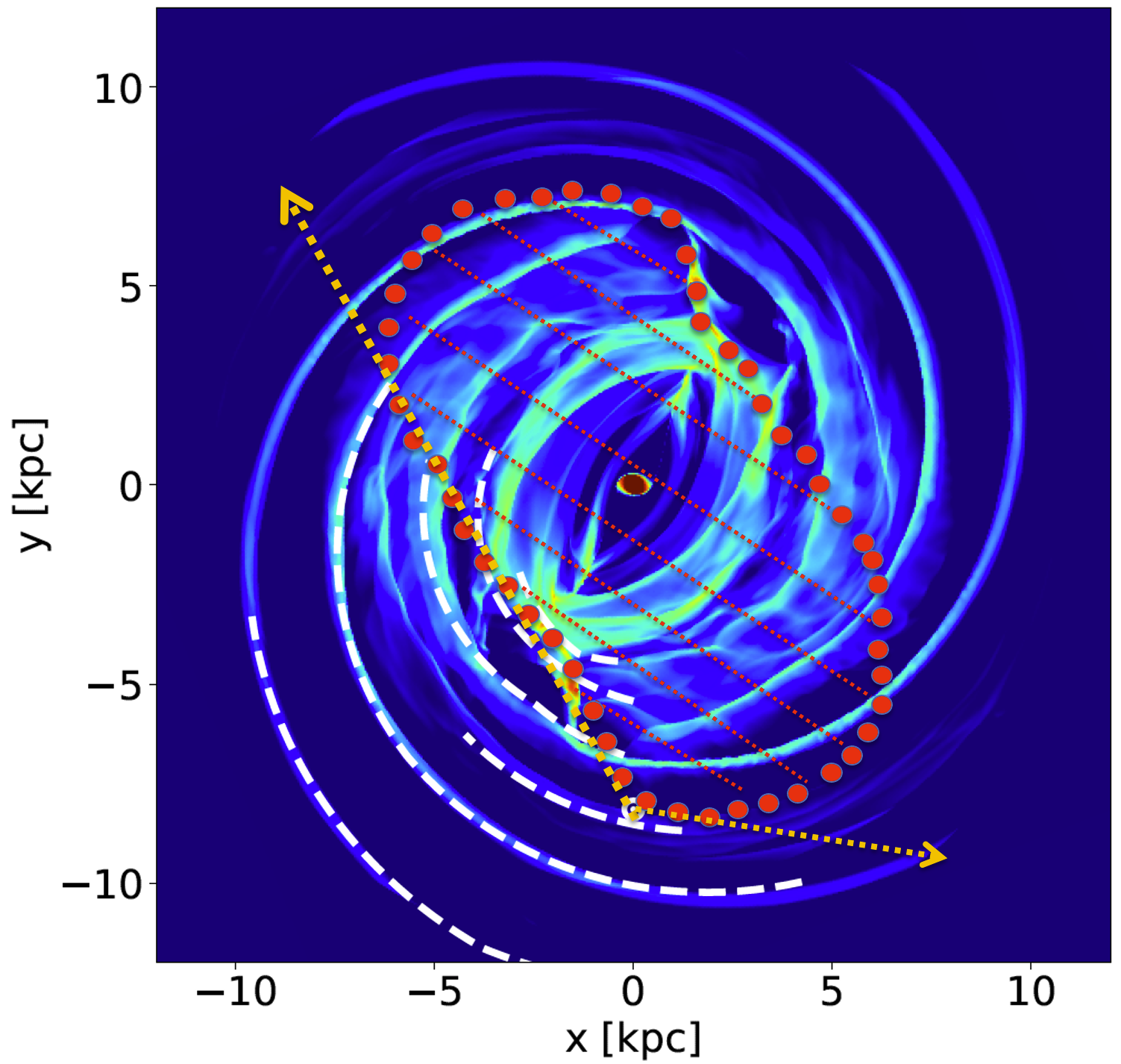}
\vspace{-6pt}
\caption{Simplistic scheme of the speculative solution discussed in this draft. The gas distribution in the disk (as viewed from above the Plane) is from the model of~\cite{Li21} already shown in Figure~\ref{fig:gasmodel}. The hypothetical boundary of the shocked halo gas generated by the FBs and having flown at short distance from the Plane is shown in projection onto the Plane (red circles). Due to the asymmetry of the gas distribution, the shock front surface has been evolving according to the obstacles that force its compression and expansion into the halo. Its expansion is such that it has now reached the Sun circle with a fully distorted boundary. As viewed from the Sun, X-rays emitted by the volume of hot gas between the FB contours and the shock front (red hashes) start abruptly from $\rm l\simeq 25$\fdeg\!\!, keep a strong brightness in the east and towards the GC where the path length is the longer, while in the west their brightness is gradually fading away down to $\rm l \simeq 270$\fdeg\!\!. The intensity of the polarized synchrotron emission is maximal in directions tangential to the shock (orange dashed arrows), i.e. mainly in the east and in a much smaller extent in the west towards $\rm l\simeq 255$\fdeg\!\!. The latter region could correspond to the western Loop I continuation proposed by~\cite{Planck16}. At high southern latitudes X-rays and synchrotron are significantly weaker than in the north due to halo weaker density.}\label{newscheme}
\end{figure}

In this article we have reviewed recent measurements and models having a direct or indirect connection with the LNPS structure and developed further comparisons, with particular focus on the use of 3D maps of interstellar gas and dust and on X-ray data. In some cases, the potential connection with the LNPS has been overlooked. Our goal is to shed additional light on the longstanding uncertainty on North Polar Spur-Loop I distance (LNPS). Several conclusions can be drawn and we summarize them below.

\vspace{-30pt}
\subsection*{}
The whole set of analyses of X-ray data, from the most recent to updated analyses of the earliest spectra (XMM-Newton, Suzaku, Swift, HaloSat) gives consistent results and favors a distant origin of the X-ray LNPS as well as an homogeneity of the spectral shapes over the whole structure. There is no firmly established upper limit on the source distance if one takes uncertainties into account, whatever the latitude range.

\begin{itemize}
\item[--] Fitted neutral gas absorbing columns are on the order of full Galactic columns at all latitudes from the lowest $\rm b=+5.6$\fdeg (\!\!\cite{Lall16}) to the highest $\rm b=+55$\fdeg (\!\!\cite{Kataoka15}). Note that, due to the dust scale height, these columns are reached at very different distances, from $\sim100$~pc at high latitude to $\sim 1$~kpc at the lowest latitudes of the measurements, see Figure~\ref{fig:absorbX}.

\item[--] At high latitude, the best fitted model spectra require a large absorbing column of hot (million K) ionized gas in addition to the dense phase (\!\!\cite{Gu16}), an evidence for a Galactic size source.

\item[--] The geometrical pattern implied by constraints on dense interstellar matter absorbing columns is hardly consistent with a nearby super bubble (Figure~\ref{fig:absorbX}).

\item[--] Anti-correlations between extinctions deduced from recent 3D maps of extinction density and ROSAT soft X-rays full-sky maps point to a distant source, from beyond 700-800~pc at low latitude to beyond 150~pc at high latitude.

\item[--] All spectral analyses point to homogeneous or slowly varying spectral shapes and subsequently continuously varying properties of the emitting hot gas.

\item[--] The spectral homogeneity is consistent with the regularity of the NPS eastern edge contour from low to high latitude, strongly suggestive of a unique structure. The regularity is particularly visible in the new e-Rosita map, because it is not affected by time-variable solar wind heliospheric and magnetospheric charge-exchange emissions, at variance with ROSAT.
\end{itemize}

The results on the X-ray source distance and homogeneity have strong implications. The homogeneity is not compatible with the dual top-bottom structure proposed by~\cite{Das20,Panopoulou21} and first suggested by~\cite{Sun15}. The argumentation of~\cite{Das20} is mostly based on a claimed contradiction between the large distance to the X-ray source at low latitude and the short distance to the source at high latitude. However, there is no actual determination of the high latitude distance limit in this work, because there is no use of absorption data, only a determination of the distance to the dust clouds, hypothesized to be equivalent to the distance to the source. On the contrary, there are contradictions between the X-ray homogeneity and the arguments presented by~\cite{Panopoulou21} and~\cite{Sun15}. In Section~\ref{topbot}, we detailed their arguments in favor of a dual structure based on depolarization and Faraday depth (\!\!\cite{Sun15}) or based on stellar light polarization and Planck-WMAP polarization (\!\!\cite{Panopoulou21}), and we come back to this issue below.

The second consequence of a distant X-ray source, (and this holds even if it is only several hundreds pc away), is an absence of link between the nearby high latitude tenuous interstellar clouds and the LPNS, because high latitude dust or gas structures filaments are unambiguously all very nearby at the periphery of the Local Cavity, as confirmed by recent 3D dust maps. As already mentioned in Section~\ref{linkgasxray}, such an absence of link is not so surprising if one makes a close inspection of the X-ray and dust maps. Importantly, this absence of link removes several difficulties associated with the hypothesis of a nearby re-heated super-bubble:
\begin{itemize}
\item[(i)] the absence of accelerated gas at the periphery of the super-bubble, since high velocity shocks are needed to explain gas heating and X-ray emission, and one would expect at least some high velocity fragments of the shell,
\item[(ii)] the absence of optical recombination lines, characteristics of SNR shell remnants. 
\item[(iii)] the required coincidence between the distance reached by the expanding, most recent shock and the distance to the nearby filaments surrounding the \emph{old} super bubble. 
\end{itemize}
For those reasons it makes more realistic the scenario presented by~\cite{Planck16} according to which hot gas and synchrotron emissions are due to the recent re-reheating of a previously blown super bubble (see e.g.~\cite{Planck16}).

\subsection*{Distance to the synchrotron source}

The most convincing arguments in favor of a nearby LNPS source come for high latitude synchrotron data, whose different aspects have been presented in Section~\ref{topbot}.
\begin{itemize}
\item[--] At $\rm b\geq30$\fdeg there is a strong similarity between the polarization angle of nearby starlight for stars beyond $\simeq$100 pc and LNPS synchrotron emission. At $\rm b\simeq 75$\fdeg the similarity is restricted to star distances between 100~pc and $\simeq 400$~pc, suggesting a source region within these limits. Consistently, at these latitudes there is a strong similarity between the polarization angle of nearby starlight and the one of the dust emission measured by Planck/WMAP.
\item[--] There is no measurable depolarization for $\rm b\geq30\fdeg$.
\item[--] At $\rm b\geq45$\fdeg the rotation measure due to the LNPS foreground is consistent with zero.
\end{itemize}
The various arguments suffer from the weakness of the dust and gas columns at high latitude and the resulting uncertainties, however the conclusions are all consistent.

\subsection*{The reheated super-bubble}

A way to reconcile X-ray and synchrotron data analyses is a reheated super-bubble centered several hundred parsecs away, as proposed by~\cite{Planck16}, and to assume that its near side at high latitude is as close as $\simeq100$-200 pc (see Figure~\ref{fig:loops_cuts_new}(III)). As we have seen, this is not favored by\linebreak O, B stars distributions, nor by 3D distribution of clouds, especially for a SB extending on both sides of the Plane. One may argue that these arguments are not quantitative, therefore not indisputable. What is indisputable is the remarkable coincidence between the huge X-ray feature produced by the SB in the north and the similarly huge X-ray feature in the south, both surrounding the northern and southern FBs respectively.

\subsection*{Geometrical arguments}

We have seen that very recent data and models as well as overlooked results bring some answers to geometrical difficulties with the hypothesis of a distant FB-connected source.
\begin{itemize}
\item[--] There is an X-ray counterpart of Loop I in the south~\cite{Predehl20}.
\item[--] There are several evidences for a higher gas density in the northern halo compared to the southern one, possibly at the origin of the weakness of the southern X-ray loop, and in different distances and shapes of outer shock fronts in the northern and southern halos.
\item[--] State-of-the-art models of the Milky Way gas confirm very strong west-east asymmetries. They may influence the hot gas expansion close to the Plane. Some of gas strong over-densities other than the eastern 3~kpc ring may correspond to the bases of arches/spurs seen by Planck/WMAP (Figure~\ref{fig:gasmodel}).
\end{itemize}

\subsection*{Conclusion and speculated scenario}

Although new observations and models in favor of the distant hypothesis are more numerous than those in favor of the local source, it remains that the full set of constraints derived from the various analyses, if entirely taken for fully granted, can not be fulfilled by any purely nearby nor purely distant model. The purely nearby configuration requires a super-bubble with a geometry and physical properties that are not favored by stellar and interstellar data and a remarkable, unlikely resemblance between the contours of the super-bubble in the north and those of the\linebreak X-ray bubble seen by Suzaku and e-Rosita in the south, both enveloping in the same way the Fermi bubbles in 2D~images. The purely distant configuration is not compatible with the estimated proximity of the synchrotron source at high latitude in the north and lacks loops in the west and in the south that could be identified with the expected huge synchrotron emitting shell.

On the other hand, several models we already mentioned are an inspiring source for a speculative scenario potentially consistent with all data. Models of FBs and associated halo heating from~\cite{Sarkar19} predict that X-rays and synchrotron shells may be different in the north and in the south due to halo density asymmetry, and that the outer shock may have reached the Sun in the south and not yet in the north. These models do not include the complex gaseous disk structure and the gap between the CMZ and the 3~kpc ring (from, e.g., \cite{Li21}). \cite{Sofue21} emphasizes the role of the ring as responsible for the low latitude spur but uses an axisymmetric halo. Combining in the same model a realistic distribution of the gas in the disk and a non-symmetric halo is a challenging task. However one may speculate that in this case the volume of shocked halo gas at low altitude above the Plane \emph{may be} very irregular, having expanded quasi-freely outside of the CMZ, then encountered beyond 3~kpc very different gas concentrations as a function of longitude. The volume of hot gas may be also very different in the southern and northern hemispheres due to different thick disk and halo gas density distribution. We have searched for a configuration able to explain the extent of the X-rays and the presence/absence of observed synchrotron shells. Figure~\ref{newscheme} shows a \emph{very} speculative outer shock contour at short distance from the Plane, distorted under the influence of the disk structures and having reached the Sun. This type of contour may potentially explain east-west differences, namely a large X-ray brightness in the east and a gradual decrease to the west, and the absence in the west of a synchrotron counterpart of the bright LPNS (see figure and caption). Evidently, such a configuration is not purely local nor purely distant but intermediate in the sense that the origin of the LNPS is distant but part of the source is nearby.

This is certainly not the end of the story and we emphasize that the suggested speculative scenario lacks modeling support. New large-scale models of the FBs and of their expansion are needed, using more complex gas distributions in the disk, like the one shown in Figure~\ref{fig:gasmodel}, and in the halo, taking into account the density asymmetry. This is a particularly difficult task, however the results may show whether or not the observed radio/microwave spurs and Loop I can be reproduced in 3D, and not only their approximate directions as in Figure~\ref{fig:gasmodel}. More sensitive and numerous radio/submm data and especially Faraday tomography studies (with LOFAR) will be also crucial to confirm or not the proximity of the high latitude Loop I. Future Gaia data releases and resulting dust maps with higher resolution and larger extent are also expected to help, and in the far future Athena X-ray spectra should bring further constraints.


\bibliographystyle{crunsrt}
\bibliography{npsbib}

\def\bysame{\leavevmode ---------\thinspace}
\makeatletter\if@francais\providecommand{\og}{<<~}\providecommand{\fg}{~>>}
\else\gdef\og{``}\gdef\fg{''}\fi\makeatother
\def\cdrandname{\&}
\providecommand\cdrnumero{no.~}
\providecommand{\cdredsname}{eds.}
\providecommand{\cdredname}{ed.}
\providecommand{\cdrchapname}{chap.}
\providecommand{\cdrmastersthesisname}{Memoir}
\providecommand{\cdrphdthesisname}{PhD Thesis}
\begin{thebibliography}{10}

\bibitem{Haslam74}
C.~G.~T. {Haslam}, W.~E. {Wilson}, D.~A. {Graham}, G.~C. {Hunt}, {\og {A
  further 408 MHz survey of the northern sky}\fg}, \emph{Astronomy \&
  Astrophysics} \textbf{13} (1974), p.~359.

\bibitem{Remazeilles15}
M.~{Remazeilles}, C.~{Dickinson}, A.~J. {Banday}, M.~A. {Bigot-Sazy},
  T.~{Ghosh}, {\og {An improved source-subtracted and destriped 408-MHz all-sky
  map}\fg}, \emph{Monthly Notices of the Royal Academy of Sciences}
  \textbf{451} (2015), \cdrnumero 4, p.~4311-4327,
  \url{https://arxiv.org/abs/1411.3628}.

\bibitem{Predehl20}
P.~{Predehl}, R.~A. {Sunyaev}, W.~{Becker}, H.~{Brunner}, R.~{Burenin},
  A.~{Bykov}, A.~{Cherepashchuk}, N.~{Chugai}, E.~{Churazov}, V.~{Doroshenko},
  N.~{Eismont}, M.~{Freyberg}, M.~{Gilfanov}, F.~{Haberl}, I.~{Khabibullin},
  R.~{Krivonos}, C.~{Maitra}, P.~{Medvedev}, A.~{Merloni}, K.~{Nandra},
  V.~{Nazarov}, M.~{Pavlinsky}, G.~{Ponti}, J.~S. {Sanders}, M.~{Sasaki},
  S.~{Sazonov}, A.~W. {Strong}, J.~{Wilms}, {\og {Detection of large-scale
  X-ray bubbles in the Milky Way halo}\fg}, \emph{Nature} \textbf{588} (2020),
  \cdrnumero 7837, p.~227-231, \url{https://arxiv.org/abs/2012.05840}.

\bibitem{Planck16}
{Planck Collaboration et al.}, {\og {Planck 2015 results. XXV. Diffuse
  low-frequency Galactic foregrounds}\fg}, \emph{Astron. \& Astrophys.}
  \textbf{594} (2016),  article no.~A25,
  \url{https://arxiv.org/abs/1506.06660}.

\bibitem{Panopoulou21}
G.~V. {Panopoulou}, C.~{Dickinson}, A.~C.~S. {Readhead}, T.~J. {Pearson}, M.~W.
  {Peel}, {\og {Revisiting the distance to radio Loops I and IV using Gaia and
  radio/optical polarization data}\fg}, \emph{arXiv e-prints} (2021),  article
  no.~arXiv:2106.14267, \url{https://arxiv.org/abs/2106.14267}.

\bibitem{Large62}
M.~I. {Large}, M.~J.~S. {Quigley}, C.~G.~T. {Haslam}, {\og {A new feature of
  the radio sky}\fg}, \emph{Monthly Notices of the Royal Academy of Sciences}
  \textbf{124} (1962), p.~405.

\bibitem{Berkhuijsen71}
E.~M. {Berkhuijsen}, {\og {A Survey of the Continuum Radiation at 820 MHz
  between Declinations -7{\textdegree} and +85{\textdegree}. A Study of the
  Galactic Radiation and the Degree of Polarization with Special Reference to
  the Loops and Spurs}\fg}, \emph{Astronomy \& Astrophysics} \textbf{14}
  (1971), p.~359.

\bibitem{SofueReich79}
Y.~{Sofue}, W.~{Reich}, {\og {Radio continuum observations of the North Polar
  Spur at 1420 MHz.}\fg}, \emph{Astronomy \& Astrophysics} \textbf{38} (1979),
  p.~251-263.

\bibitem{Snowden97}
S.~L. {Snowden}, R.~{Egger}, M.~J. {Freyberg}, D.~{McCammon}, P.~P.
  {Plucinsky}, W.~T. {Sanders}, J.~H.~M.~M. {Schmitt}, J.~{Tr{\"u}mper},
  W.~{Voges}, {\og {ROSAT Survey Diffuse X-Ray Background Maps. II.}\fg},
  \emph{Astrophysical Journal} \textbf{485} (1997), \cdrnumero 1, p.~125-135.

\bibitem{Bingham67}
R.~G. {Bingham}, {\og {Magnetic fields in the galactic spurs}\fg},
  \emph{Monthly Notices of the Royal Academy of Sciences} \textbf{137} (1967),
  p.~157.

\bibitem{Brouw76}
W.~N. {Brouw}, T.~A.~T. {Spoelstra}, {\og {Linear polarization of the galactic
  background at frequencies between 408 and 1411 MHz. Reductions.}\fg},
  \emph{Astronomy \& Astrophysics} \textbf{26} (1976), p.~129.

\bibitem{Casandjian09}
J.-M. {Casandjian}, I.~{Grenier}, {\og {High Energy Gamma-Ray Emission from the
  Loop I region}\fg}, \emph{arXiv e-prints} (2009),  article
  no.~arXiv:0912.3478, \url{https://arxiv.org/abs/0912.3478}.

\bibitem{Sofue77}
Y.~{Sofue}, {\og {Propagation of magnetohydrodynamic waves from the galactic
  center. Origin of the 3-kpc arm and the North Polar Spur.}\fg}, \emph{Astron.
  \& Astrophys.} \textbf{60} (1977), \cdrnumero 3, p.~327-336.

\bibitem{Finkbeiner04}
D.~P. {Finkbeiner}, {\og {Microwave Interstellar Medium Emission Observed by
  the Wilkinson Microwave Anisotropy Probe}\fg}, \emph{Astrophysical Journal}
  \textbf{614} (2004), \cdrnumero 1, p.~186-193,
  \url{https://arxiv.org/abs/astro-ph/0311547}.

\bibitem{Dobler08}
G.~{Dobler}, D.~P. {Finkbeiner}, {\og {Extended Anomalous Foreground Emission
  in the WMAP Three-Year Data}\fg}, \emph{Astrophysical Journal} \textbf{680}
  (2008), \cdrnumero 2, p.~1222-1234, \url{https://arxiv.org/abs/0712.1038}.

\bibitem{Dobler10}
G.~{Dobler}, D.~P. {Finkbeiner}, I.~{Cholis}, T.~{Slatyer}, N.~{Weiner}, {\og
  {The Fermi Haze: A Gamma-ray Counterpart to the Microwave Haze}\fg},
  \emph{Astrophysical Journal} \textbf{717} (2010), \cdrnumero 2, p.~825-842,
  \url{https://arxiv.org/abs/0910.4583}.

\bibitem{Su10}
M.~{Su}, T.~R. {Slatyer}, D.~P. {Finkbeiner}, {\og {Giant Gamma-ray Bubbles
  from Fermi-LAT: Active Galactic Nucleus Activity or Bipolar Galactic
  Wind?}\fg}, \emph{Astrophysical Journal} \textbf{724} (2010), \cdrnumero 2,
  p.~1044-1082, \url{https://arxiv.org/abs/1005.5480}.

\bibitem{Crocker15}
R.~M. {Crocker}, G.~V. {Bicknell}, A.~M. {Taylor}, E.~{Carretti}, {\og {A
  Unified Model of the Fermi Bubbles, Microwave Haze, and Polarized Radio
  Lobes: Reverse Shocks in the Galactic Center{\textquoteright}s Giant
  Outflows}\fg}, \emph{Astrophys. J.} \textbf{808} (2015), \cdrnumero 2,
  article no.~107, \url{https://arxiv.org/abs/1412.7510}.

\bibitem{Sarkar15}
K.~C. {Sarkar}, B.~B. {Nath}, P.~{Sharma}, {\og {Multiwavelength features of
  Fermi bubbles as signatures of a Galactic wind}\fg}, \emph{Monthly Notices of
  the Royal Academy of Sciences} \textbf{453} (2015), \cdrnumero 4,
  p.~3827-3838, \url{https://arxiv.org/abs/1505.03634}.

\bibitem{Sarkar19}
K.~C. {Sarkar}, {\og {Possible connection between the asymmetry of the North
  Polar Spur and Loop I and Fermi bubbles}\fg}, \emph{MNRAS} \textbf{482}
  (2019), \cdrnumero 4, p.~4813-4823, \url{https://arxiv.org/abs/1804.05634}.

\bibitem{Ponti21}
G.~{Ponti}, M.~R. {Morris}, E.~{Churazov}, I.~{Heywood}, R.~P. {Fender}, {\og
  {The Galactic center chimneys: the base of the multiphase outflow of the
  Milky Way}\fg}, \emph{Astronomy \& Astrophysics} \textbf{646} (2021),
  article no.~A66, \url{https://arxiv.org/abs/2101.05284}.

\bibitem{Cecil21}
G.~{Cecil}, A.~Y. {Wagner}, J.~{Bland-Hawthorn}, G.~V. {Bicknell},
  D.~{Mukherjee}, {\og {Tracing the Milky Way's Vestigial Nuclear Jet}\fg},
  \emph{arXiv e-prints} (2021),  article no.~arXiv:2109.00901,
  \url{https://arxiv.org/abs/2109.00901}.

\bibitem{Casandjian15}
J.-M. {Casandjian}, {\og {The Fermi-LAT model of interstellar emission for
  standard point source analysis}\fg}, \emph{arXiv e-prints} (2015),  article
  no.~arXiv:1502.07210, \url{https://arxiv.org/abs/1502.07210}.

\bibitem{Vergely22}
J.~{Vergely}, R.~{Lallement}, N.~{Cox}, {\og {3D extinction maps: inversion of
  inter-calibrated extinction catalogs}\fg}, \emph{in prep.} (2022).

\bibitem{Puspitarini12}
L.~{Puspitarini}, R.~{Lallement}, {\og {Distance to northern high-latitude HI
  shells}\fg}, \emph{Astron. \& Astrophys.} \textbf{545} (2012),  article
  no.~A21, \url{https://arxiv.org/abs/1207.5353}.

\bibitem{Tahara15}
M.~{Tahara}, J.~{Kataoka}, Y.~{Takeuchi}, T.~{Totani}, Y.~{Sofue}, J.~S.
  {Hiraga}, H.~{Tsunemi}, Y.~{Inoue}, M.~{Kimura}, C.~C. {Cheung},
  S.~{Nakashima}, {\og {Suzaku X-Ray Observations of the Fermi Bubbles:
  Northernmost Cap and Southeast Claw Discovered With MAXI-SSC}\fg},
  \emph{Astrophysical Journal} \textbf{802} (2015), \cdrnumero 2,  article
  no.~91, \url{https://arxiv.org/abs/1501.04405}.

\bibitem{Vidal15}
M.~{Vidal}, C.~{Dickinson}, R.~D. {Davies}, J.~P. {Leahy}, {\og {Polarized
  radio filaments outside the Galactic plane}\fg}, \emph{Monthly Notices of the
  Royal Academy of Sciences} \textbf{452} (2015), \cdrnumero 1, p.~656-675,
  \url{https://arxiv.org/abs/1410.4438}.

\bibitem{French21}
D.~M. {French}, A.~J. {Fox}, B.~P. {Wakker}, C.~{Norman}, N.~{Lehner}, J.~C.
  {Howk}, B.~D. {Savage}, P.~{Richter}, J.~{O'Meara}, S.~{Borthakur},
  T.~{Heckman}, {\og {The HI Column Density Distribution of the Galactic Disk
  and Halo}\fg}, \emph{arXiv e-prints} (2021),  article no.~arXiv:2108.07419,
  \url{https://arxiv.org/abs/2108.07419}.

\bibitem{Qu20}
Z.~{Qu}, J.~N. {Bregman}, E.~{Hodges-Kluck}, J.-T. {Li}, R.~{Lindley}, {\og
  {The Warm Gas in the MW: A Kinematical Model}\fg}, \emph{Astrophys. J.}
  \textbf{894} (2020), \cdrnumero 2,  article no.~142,
  \url{https://arxiv.org/abs/2002.06434}.

\bibitem{Ashley20}
T.~{Ashley}, A.~J. {Fox}, E.~B. {Jenkins}, B.~P. {Wakker}, R.~{Bordoloi}, F.~J.
  {Lockman}, B.~D. {Savage}, T.~{Karim}, {\og {Mapping Outflowing Gas in the
  Fermi Bubbles: A UV Absorption Survey of the Galactic Nuclear Wind}\fg},
  \emph{Astrophysical Journal} \textbf{898} (2020), \cdrnumero 2,  article
  no.~128, \url{https://arxiv.org/abs/2006.13254}.

\bibitem{Li21}
Z.~{Li}, J.~{Shen}, O.~{Gerhard}, J.~P. {Clarke}, {\og {Gas Dynamics in the
  Galaxy: Total Mass Distribution and the Bar Pattern Speed}\fg}, \emph{arXiv
  e-prints} (2021),  article no.~arXiv:2103.10342,
  \url{https://arxiv.org/abs/2103.10342}.

\bibitem{Sofue21}
Y.~{Sofue}, J.~{Kataoka}, {\og {Interaction of the galactic-centre super
  bubbles with the gaseous disc}\fg}, \emph{MNRAS} \textbf{506} (2021),
  \cdrnumero 2, p.~2170-2180, \url{https://arxiv.org/abs/2106.14955}.

\bibitem{Wolleben06}
M.~{Wolleben}, T.~L. {Landecker}, W.~{Reich}, R.~{Wielebinski}, {\og {An
  absolutely calibrated survey of polarized emission from the northern sky at
  1.4 GHz. Observations and data reduction}\fg}, \emph{Astronomy and
  Astrophysics} \textbf{448} (2006), \cdrnumero 1, p.~411-424,
  \url{https://arxiv.org/abs/astro-ph/0510456}.

\bibitem{Dickinson18}
C.~{Dickinson}, {\og {Large-Scale Features of the Radio Sky and a Model for
  Loop I}\fg}, \emph{Galaxies} \textbf{6} (2018), \cdrnumero 2, p.~56.

\bibitem{Sun15}
X.~H. {Sun}, T.~L. {Landecker}, B.~M. {Gaensler}, E.~{Carretti}, W.~{Reich},
  J.~P. {Leahy}, N.~M. {McClure-Griffiths}, R.~M. {Crocker}, M.~{Wolleben},
  M.~{Haverkorn}, K.~A. {Douglas}, A.~D. {Gray}, {\og {Faraday Tomography of
  the North Polar Spur: Constraints on the Distance to the Spur and on the
  Magnetic Field of the Galaxy}\fg}, \emph{Astrophysical Journal} \textbf{811}
  (2015), \cdrnumero 1,  article no.~40,
  \url{https://arxiv.org/abs/1508.03889}.

\bibitem{Oppermann15}
N.~{Oppermann}, H.~{Junklewitz}, M.~{Greiner}, T.~A. {En{\ss}lin},
  T.~{Akahori}, E.~{Carretti}, B.~M. {Gaensler}, A.~{Goobar},
  L.~{Harvey-Smith}, M.~{Johnston-Hollitt}, L.~{Pratley}, D.~H.~F.~M.
  {Schnitzeler}, J.~M. {Stil}, V.~{Vacca}, {\og {Estimating extragalactic
  Faraday rotation}\fg}, \emph{Astronomy and Astrophysics} \textbf{575} (2015),
   article no.~A118, \url{https://arxiv.org/abs/1404.3701}.

\bibitem{Akita18}
M.~{Akita}, J.~{Kataoka}, M.~{Arimoto}, Y.~{Sofue}, T.~{Totani}, Y.~{Inoue},
  S.~{Nakashima}, {\og {Diffuse X-Ray Emission from the Northern Arc of Loop I
  Observed with Suzaku}\fg}, \emph{Astrophysical Journal} \textbf{862} (2018),
  \cdrnumero 1,  article no.~88, \url{https://arxiv.org/abs/1806.08058}.

\bibitem{Kataoka13}
J.~{Kataoka}, M.~{Tahara}, T.~{Totani}, Y.~{Sofue}, {\L}.~{Stawarz},
  Y.~{Takahashi}, Y.~{Takeuchi}, H.~{Tsunemi}, M.~{Kimura}, Y.~{Takei}, C.~C.
  {Cheung}, Y.~{Inoue}, T.~{Nakamori}, {\og {Suzaku Observations of the Diffuse
  X-Ray Emission across the Fermi Bubbles' Edges}\fg}, \emph{Astrophysical
  Journal} \textbf{779} (2013), \cdrnumero 1,  article no.~57,
  \url{https://arxiv.org/abs/1310.3553}.

\bibitem{Kataoka15}
J.~{Kataoka}, M.~{Tahara}, T.~{Totani}, Y.~{Sofue}, Y.~{Inoue}, S.~{Nakashima},
  C.~C. {Cheung}, {\og {Global Structure of Isothermal Diffuse X-Ray Emission
  along the Fermi Bubbles}\fg}, \emph{Astrophysical Journal} \textbf{807}
  (2015), \cdrnumero 1,  article no.~77,
  \url{https://arxiv.org/abs/1505.05936}.

\bibitem{Kaaret19}
P.~{Kaaret}, A.~{Zajczyk}, D.~M. {LaRocca}, R.~{Ringuette}, J.~{Bluem},
  W.~{Fuelberth}, H.~{Gulick}, K.~{Jahoda}, T.~E. {Johnson}, D.~L. {Kirchner},
  D.~{Koutroumpa}, K.~D. {Kuntz}, R.~{McCurdy}, D.~M. {Miles}, W.~T. {Robison},
  E.~M. {Silich}, {\og {HaloSat: A CubeSat to Study the Hot Galactic Halo}\fg},
  \emph{Astrophysical Journal} \textbf{884} (2019), \cdrnumero 2,  article
  no.~162, \url{https://arxiv.org/abs/1909.13822}.

\bibitem{LaRocca20}
D.~M. {LaRocca}, P.~{Kaaret}, K.~D. {Kuntz}, E.~{Hodges-Kluck}, A.~{Zajczyk},
  J.~{Bluem}, R.~{Ringuette}, K.~M. {Jahoda}, {\og {An Analysis of the North
  Polar Spur Using HaloSat}\fg}, \emph{Astrophys. J.} \textbf{904} (2020),
  \cdrnumero 1,  article no.~54.

\bibitem{Lall16}
R.~{Lallement}, S.~{Snowden}, K.~D. {Kuntz}, T.~M. {Dame}, D.~{Koutroumpa},
  I.~{Grenier}, J.~M. {Casandjian}, {\og {On the distance to the North Polar
  Spur and the local CO-H$_{2}$ factor}\fg}, \emph{Astron. \& Astrophys.}
  \textbf{595} (2016),  article no.~A131,
  \url{https://arxiv.org/abs/1609.03813}.

\bibitem{Willingale03}
R.~{Willingale}, A.~D.~P. {Hands}, R.~S. {Warwick}, S.~L. {Snowden}, D.~N.
  {Burrows}, {\og {The X-ray spectrum of the North Polar Spur}\fg},
  \emph{MNRAS} \textbf{343} (2003), \cdrnumero 3, p.~995-1001.

\bibitem{Miller08}
E.~D. {Miller}, {Tsunemi Hiroshi}, M.~W. {Bautz}, D.~{McCammon}, R.~{Fujimoto},
  J.~P. {Hughes}, S.~{Katsuda}, M.~{Kokubun}, K.~{Mitsuda}, F.~S. {Porter},
  Y.~{Takei}, Y.~{Tsuboi}, N.~Y. {Yamasaki}, {\og {Suzaku Observations of the
  North Polar Spur: Evidence for Nitrogen Enhancement}\fg}, \emph{Publications
  of the Astronomical Society of Japan} \textbf{60} (2008), p.~S95,
  \url{https://arxiv.org/abs/0708.4227}.

\bibitem{Snowden95}
S.~L. {Snowden}, M.~J. {Freyberg}, P.~P. {Plucinsky}, J.~H.~M.~M. {Schmitt},
  J.~{Truemper}, W.~{Voges}, R.~J. {Edgar}, D.~{McCammon}, W.~T. {Sanders},
  {\og {First Maps of the Soft X-Ray Diffuse Background from the ROSAT XRT/PSPC
  All-Sky Survey}\fg}, \emph{Astrophysical Journal} \textbf{454} (1995),
  p.~643.

\bibitem{Das20}
K.~K. {Das}, C.~{Zucker}, J.~S. {Speagle}, A.~{Goodman}, G.~M. {Green},
  J.~{Alves}, {\og {Constraining the distance to the North Polar Spur with Gaia
  DR2}\fg}, \emph{MNRAS} \textbf{498} (2020), \cdrnumero 4, p.~5863-5872,
  \url{https://arxiv.org/abs/2009.01320}.

\bibitem{Lall09}
R.~{Lallement}, {\og {Some Observations Related to the Origin and Evolution of
  the Local Bubble/Local ISM}\fg}, \emph{Space Science Reviews} \textbf{143}
  (2009), \cdrnumero 1-4, p.~427-436.

\bibitem{Gu16}
L.~{Gu}, J.~{Mao}, E.~{Costantini}, J.~{Kaastra}, {\og {Suzaku and XMM-Newton
  observations of the North Polar Spur: Charge exchange or ISM
  absorption?}\fg}, \emph{Astronomy \& Astrophysics} \textbf{594} (2016),
  article no.~A78, \url{https://arxiv.org/abs/1607.08334}.

\bibitem{MacLow89}
M.-M. {Mac Low}, R.~{McCray}, M.~L. {Norman}, {\og {Superbubble Blowout
  Dynamics}\fg}, \emph{Astrophysical Journal} \textbf{337} (1989), p.~141.

\bibitem{Puspitarini14}
L.~{Puspitarini}, R.~{Lallement}, J.~L. {Vergely}, S.~L. {Snowden}, {\og {Local
  ISM 3D distribution and soft X-ray background. Inferences on nearby hot gas
  and the North Polar Spur}\fg}, \emph{Astron. \& Astrophys.} \textbf{566}
  (2014),  article no.~A13, \url{https://arxiv.org/abs/1401.6899}.

\bibitem{Wolleben07}
M.~{Wolleben}, {\og {A New Model for the Loop I (North Polar Spur) Region}\fg},
  \emph{Astrophysical Journal} \textbf{664} (2007), \cdrnumero 1, p.~349-356,
  \url{https://arxiv.org/abs/0704.0276}.

\bibitem{Breitschwerdt16}
D.~{Breitschwerdt}, J.~{Feige}, M.~M. {Schulreich}, M.~A.~D. {Avillez},
  C.~{Dettbarn}, B.~{Fuchs}, {\og {The locations of recent supernovae near the
  Sun from modelling $^{60}$Fe transport}\fg}, \emph{Nature} \textbf{532}
  (2016), \cdrnumero 7597, p.~73-76.

\bibitem{DeAvillez05}
M.~A. {de Avillez}, D.~{Breitschwerdt}, {\og {Global dynamical evolution of the
  ISM in star forming galaxies. I. High resolution 3D simulations: Effect of
  the magnetic field}\fg}, \emph{Astronomy and Astrophysics} \textbf{436}
  (2005), \cdrnumero 2, p.~585-600,
  \url{https://arxiv.org/abs/astro-ph/0502327}.

\bibitem{Pantaleoni21}
M.~{Pantaleoni Gonz{\'a}lez}, J.~{Ma{\'\i}z Apell{\'a}niz}, R.~H. {Barb{\'a}},
  B.~C. {Reed}, {\og {The Alma catalogue of OB stars - II. A cross-match with
  Gaia DR2 and an updated map of the solar neighbourhood}\fg}, \emph{Monthly
  Notices of the Royal Academy of Sciences} \textbf{504} (2021), \cdrnumero 2,
  p.~2968-2982, \url{https://arxiv.org/abs/2103.02748}.

\bibitem{Robitaille18}
J.~F. {Robitaille}, A.~M.~M. {Scaife}, E.~{Carretti}, M.~{Haverkorn}, R.~M.
  {Crocker}, M.~J. {Kesteven}, S.~{Poppi}, L.~{Staveley-Smith}, {\og
  {Interstellar magnetic cannon targeting the Galactic halo. A young bubble at
  the origin of the Ophiuchus and Lupus molecular complexes}\fg},
  \emph{Astronomy and Astrophysics} \textbf{617} (2018),  article no.~A101,
  \url{https://arxiv.org/abs/1807.04054}.

\bibitem{Lall19}
R.~{Lallement}, C.~{Babusiaux}, J.~L. {Vergely}, D.~{Katz}, F.~{Arenou},
  B.~{Valette}, C.~{Hottier}, L.~{Capitanio}, {\og {Gaia-2MASS 3D maps of
  Galactic interstellar dust within 3 kpc}\fg}, \emph{Astronomy and
  Astrophysics} \textbf{625} (2019),  article no.~A135,
  \url{https://arxiv.org/abs/1902.04116}.

\end{thebibliography}
\end{document}